\documentclass[twocolumn,showpacs,aps,amsmath,amssymb,superscriptaddress,prb,citeautoscript]{revtex4-1}
\usepackage{amsfonts}
\usepackage{amsmath}
\usepackage{bm}
\usepackage{setspace}
\usepackage{graphicx}
\usepackage{color}
\usepackage{multirow}
\usepackage{verbatim}
\usepackage{float}
\usepackage{booktabs}
\usepackage[normalem]{ulem}

\newcommand{\nl}{\nonumber \\}

\newcommand{\be}{\begin{equation}}
\newcommand{\ee}{\end{equation}}
\newcommand{\bea}{\begin{eqnarray}}
\newcommand{\eea}{\end{eqnarray}}
\newcommand{\bsube}{\begin{subequations}}
\newcommand{\esube}{\end{subequations}}
\newcommand{\Eq}[1]{Eq.~(\ref{#1})}

\newcommand{\Fig}[1]{Fig.~\ref{#1}}

\newcommand{\Tab}[1]{Table~\ref{#1}}

\newcommand{\br}{\mathbf{r}}

\newcommand{\openbox}{\leavevmode
  \hbox to.77778em{%
  \hfil\vrule
  \vbox to.675em{\hrule width.6em\vfil\hrule}%
  \vrule\hfil}}

\newcommand{\mO}{\mathcal{O}}
\newcommand{\heff}{\hbar_{\text{eff}}}

\begin{document}
	
\renewcommand{\figurename}{Figure}
\renewcommand{\tablename}{Table}
\renewcommand{\thetable}{\arabic{table}}

\title{Exact Constraint of Density Functional Approximations \\ at the Semiclassical Limit}

\author{Yunzhi Li} \author{Chen Li} \email{chenlichem@pku.edu.cn}
\affiliation{Beijing National Laboratory for Molecular Sciences, College of Chemistry and Molecular Engineering, Peking University,
Beijing 100871, China}

\date{\today}

\begin{abstract}

{\bf ABSTRACT:}
We introduce the semiclassical limit to electronic systems by taking the limit $\hbar\rightarrow 0$ in the solution of Schr\"odinger equations. We show that this limit is closely related to one type of strong correlation that is particularly challenging from conventional multi-configurational perspective but can be readily described through semiclassical analysis. 
Furthermore, by studying the performance of density functional approximations (DFAs) in the semiclassical limit, we find that mainstream DFAs have erroneous divergent energy behaviors as $\hbar \rightarrow 0$, violating the exact constraint of finite energy.
Importantly, by making connection of the significantly underestimated DFA energies of many strongly correlated transition-metal diatomic molecules to their rather small estimated $\heff$, we demonstrate the usefulness of our semiclassical analysis and its promise for inspiring better DFAs.

\end{abstract}

\maketitle

\section{Introduction}

Density functional theory (DFT) with the existing approximations has been extremely successful in describing weakly correlated systems, ranging from atoms, molecules to metallic solids. Through introducing an auxiliary non-interacting system,\cite{Kohn65A1133} Kohn-Sham DFT is endowed with the physical picture of independent electrons moving in a mean field, which is essentially valid for most of the main-group compounds of chemical interests. In such a way, the complicated many-body wave function is effectively reduced to $N$ occupied orbitals. This crucial simplification makes life so much easier in terms of both theory and computation that chemists cannot live without the concept of orbitals since the early days of quantum mechanics until the present day.
However, there is one class of challenging systems where the single-electron orbital picture breaks down, known as strongly correlated systems.
Typical examples of strongly correlated systems include stretched molecules and transition-metal compounds, \cite{Chan193610,Cohen08792,Kulik100021}
where the correlation cannot be captured by perturbative treatment to any single Slater determinant, and one cannot find a dominant determinant in the configuration interaction (CI) expansion of the many-body wave function. \cite{Cohen12289,Becke1418A301,Mardirossian172315,Bulik153171}
Strong correlation can also occur in solid materials, such as Mott insulators, \cite{Mott3772,Mott68677,Imada981039,Boer3759,Zaanen85418,Kulik150021}
where
the strong electron--electron interaction leads to the breakdown of the band structure theory.\cite{Kotliar0453}

It is crucial to note that the above-mentioned challenges should not be regarded as failures of density functional theory, which by the Hohenberg-Kohn formulation \cite{Hohenberg64B864} is formally exact. Instead, problem arises in the exchange-correlation (xc) functional in the Kohn-Sham decomposition of the total energy, whose exact form cannot be obtained explicitly and in practical calculations has to be treated by density functional approximations (DFAs).
It is the qualitatively incorrect behaviors of these DFAs, characterized by the violation of some important conditions or constraints, that are responsible for their failures in strongly correlated systems.
Among other exact constraints, one important constraint relevant for strong correlation is the fractional spin (FS) condition. \cite{Cohen080021,Cohen08792}
In particular, the strong correlation problem of a stretched molecule is understood from the perspective that local spin densities of an isolated fragment integrate to a fractional number, a problematic scenario where DFAs tend to overestimate the energy.
Along this line of thinking, functionals have been constructed to restore the local FS condition, which can then properly describe molecular dissociation. \cite{Su189678}
However, for unstretched molecules such as well-bonded metal oxides, how one should perform the FS analysis remains a major challenge.

\begin{figure}[htpb]
	\begin{center}
		\includegraphics[width=\linewidth]{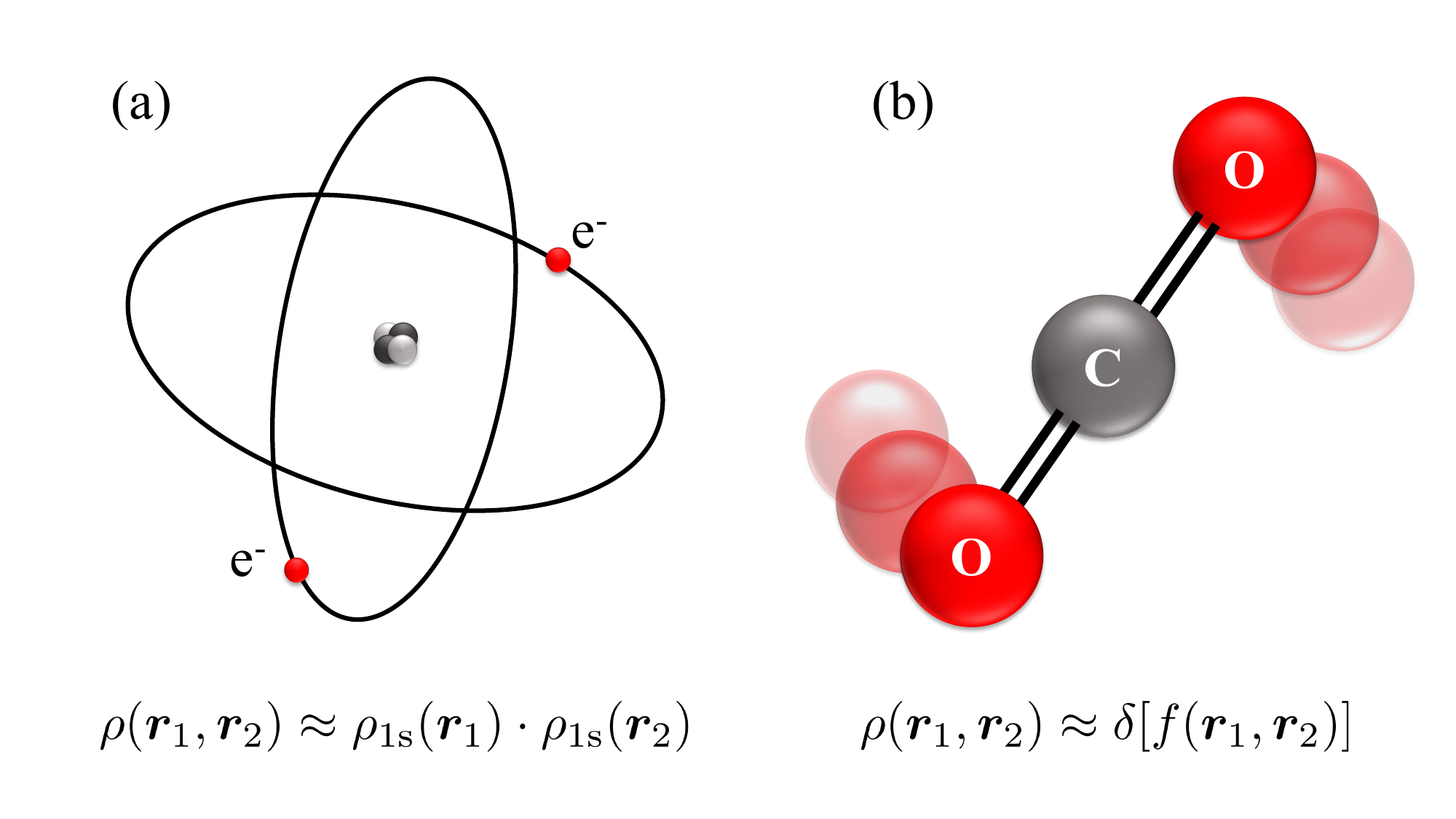}
	\end{center}
	\caption{
		Schematic illustration of the weakly and strongly correlated picture:
		(a) two electrons independently orbit around the helium nucleus, representative of a weakly correlated picture;
		(b) two oxygen atoms rotate around the carbon atom in collective motion in a CO$_2$ molecule, representative of a strongly correlated picture.
	} \label{schemWCvsSC}
\end{figure}

To gain more insight into the problem of strong correlation, let us take one step back and review the physical nature of weak and strong correlation through an example of a two-particle system.
In a weakly correlated picture such as the electrons in a helium atom in its ground state, the electronic motions are almost independent from each other, pictorially shown as in \Fig{schemWCvsSC}(a). It follows from the probability theory that the joint probability of finding one electron at $\br_1$ and the other at $\br_2$ is given by the following product, $\rho(\br_1, \br_2) \approx \rho_{1s}(\br_1)\rho_{1s}(\br_2)$, where $\rho_{1s}(\br)$ is the probability of finding an electron in the 1s orbital.
In the case of strong correlation, in contrast, the joint probability $\rho(\br_1, \br_2)$ becomes a $\delta$ function. This is because the motions of the two electrons are strongly entangled; if one knows the position of one particle, then one immediately knows the position of the other. 
Interpreted physically, having motions of collective rather than of individual nature, is what distinguishes strong correlation from weak correlation.
Such a picture is reminiscent of atoms in a molecule within the semiclassical treatment.
We can remind ourselves of the linear CO$_2$ molecule as shown in \Fig{schemWCvsSC}(b), where the two oxygen atoms are strongly entangled as they always appear equidistantly on the opposite sides of the carbon atom.

\begin{figure}[htpb]
	\begin{center}
		\includegraphics[width=\linewidth]{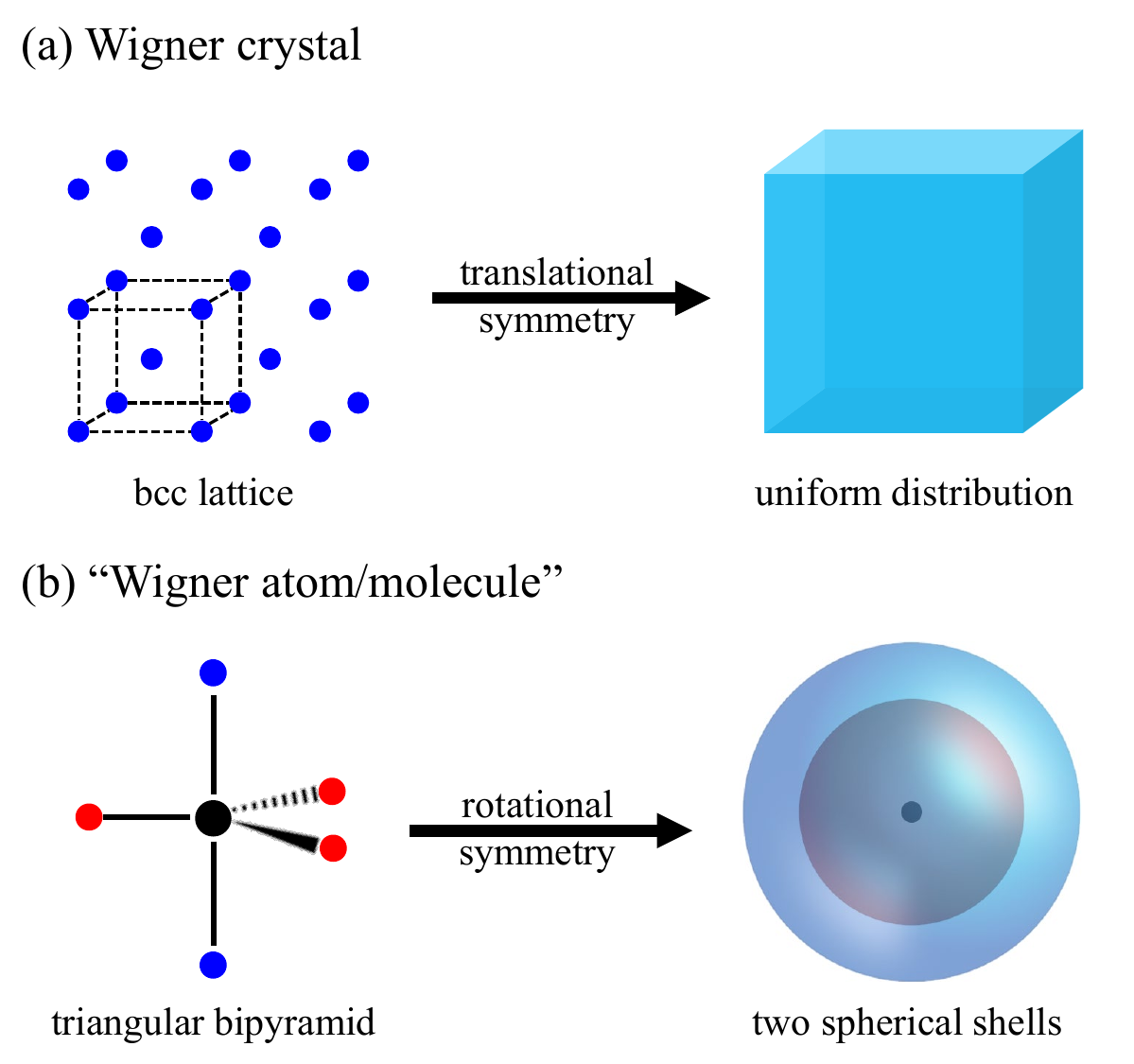}
	\end{center}
	\caption{
		Schematic illustration comparing (a) Wigner crystal and (b) a 5-electron ``Wigner atom''. In (b), electrons crystalize at two types of classical positions surrounding the nucleus (in black), as shown by the red and blue dots, respectively. After considering the rotational symmetry, they form two spherical shells.
	} \label{schemWignerCrystal}
\end{figure}

The fact that the strong correlation limit and the semiclassical limit has a lot in common is not a coincidence. Here let us remind ourselves that in the most general sense semiclassical limit is defined by taking a small $\hbar$-related dimensionless quantity to zero;\cite{Landsman07417} this quantity can be $\hbar$ itself or a combination of $\hbar$ and other parameters such as the nuclear mass. In CO$_2$, because the nuclear kinetic energy $T_n$ is much smaller than the potential energy $V$ (due to the much heavier nuclear mass compared to the electronic mass), as a good approximation by Born and Oppenheimer,\cite{Born27457} we can neglect $T_n$ and treat the nuclei as clamped at their classical positions.
This forms a major contrast with the electrons, for which the kinetic energy $T$ is for most cases on the same order of magnitude as $V$. However, there are exceptions. A famous example is the uniform electron gas (UEG) in the low density limit, where $V$ dominates over $T$ such that electrons behave like semiclassical particles that crystalize to form a lattice, known as the Wigner crystal. \cite{Wigner341002,Wigner38678,Carr611437}
\Fig{schemWignerCrystal} (a) is an illustration of the Wigner crystal for a body-centered cubic (bcc) lattice.
In the design of the simplest local density approximation (LDA) to the correlation energy by Vosko, Wilk and Nusair (VWN5) \cite{Ceperley80566,Vosko801200}, the correct limiting behaviors for low as well as high density UEG have already been built in the functional. In this sense, it comes as no surprise that LDA can properly describe metallic solids even though some metals are presumably strongly correlated- such systems are not so distinct from the UEG.

For atoms and molecules, however, the idealization of UEG or Wigner crystal barely occurs because the electron density is usually far from uniform distribution.
Nevertheless, we can still borrow the idea we learned from Wigner crystal and define a semiclassical limit, which we refer to as the ``Wigner atom/molecule'' highlighting the non-uniform density distribution and finite boundary condition.
Here we achieve this by taking the limit $\hbar \rightarrow 0$ to the solution of the Schr\"odinger equation, which is essentially to make $T$ negligibly small compared to $V$.
A schematic illustration is given in \Fig{schemWignerCrystal} (b), where five electrons condense into classical positions [electron density becomes a sum of delta functions, see \Eq{SC-rho}] and form a triangular bipyramid structure as $\hbar \rightarrow 0$, in order to reach the minimum potential energy.
When rotational symmetry is considered, the electron density shall evenly distribute on two spherical shells, with the radii corresponding to the two electron-nuclear distances of the triangular bipyramid.

Through this definition, we can study the exact constraint of DFAs at the semiclassical limit. We will show that mainstream DFAs all suffer from qualitatively wrong asymptotic behavior in approaching this limit. Moreover, by estimating the effective $\hbar$ for real molecules, we establish a semi-quantitative relation between the error of LDA [and also generalized gradient approximations (GGAs)] and $\hbar_{\rm eff}$. In particular, we find that there is a strong correlation between the large error of DFAs and small $\hbar_{\rm eff}$, particularly for transition metal compounds.
Before we go into details of our theory, it is worth mentioning the strictly correlated electron (SCE) theory developed by Seidl and Gori-Giorgi et al,\cite{Vuckovic153153,Vuckovic23e1634,Seidl07042511,Seidl994387,Gori-Giorgi09743,Buttazzo12062502}
which also discusses intensively the exact functional in the strong correlation limit for molecular systems.
Differing from the present work that focuses on the exact conditions violated by DFAs for $v$-representable densities, the SCE theory focuses on the exact functional form defined for all $N$-representable densities through the constrained search formula.
The resulting form of the SCE functional involves complicated co-motion functions, which without further approximations poses challenges for practical calculations beyond a few electrons. \cite{Vuckovic23e1634}
In this paper, we take a different route. By identifying the missing exact semiclassical constraints important for strongly correlated molecules, we hope to guide functional approximations that improve mainstream DFAs for these systems by recovering the correct limit.

\section{Theory}

\subsection{Effective $\hbar$ for real molecules}
Let us start by considering the following many-electron Hamiltonian under an external one-body potential $V_{\rm ext}$,
\begin{align}
	\Big[ -\frac{1}{2}\sum_{i}\nabla_i^2 + \sum_{i}V_{\rm ext}(\br_i) + \sum_{i<j}\frac{1}{|\br_i-\br_j|} \Big]\Psi = E\Psi, \label{SE1}
\end{align}
At first glance, it seems that there is no playground to introduce an adjustable $\hbar$ or take the semiclassical limit, because $\hbar $ is a physical constant and set to 1 in atomic unit for real molecules. In the following, we show that an effective $\hbar$ can be introduced in two ways.

\begin{figure}[htpb]
	\begin{center}
		\includegraphics[width=\linewidth]{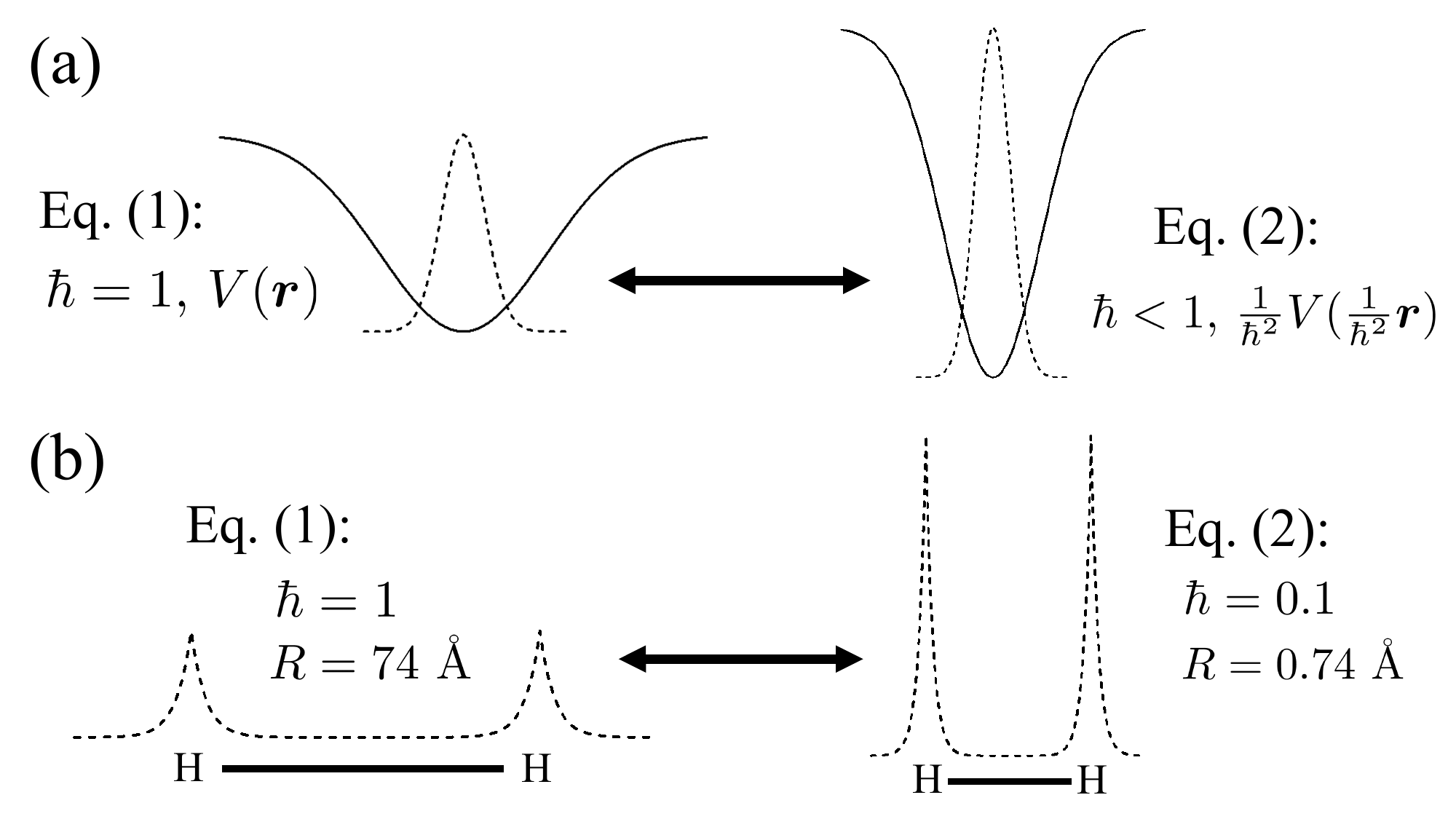}
	\end{center}
	\caption{
		Visualization of the connection between the two systems defined by \Eq{SE1} and \Eq{SEhbar}.
		Here (a) is a schematic illustration of a general problem, where the solid lines represent the external potentials and dashed lines represent the ground state densities; (b) gives a concrete example of a dissociated H$_2$ molecule (only densities are shown).
	} \label{schemScaling}
\end{figure}

The first way is through scaling the coordinates by $\br_i' \equiv \hbar^2 \br_i$ such that
we can introduce $\hbar \neq 1$ as an adjustable parameter lesser than 1. The resulting SE reads
\begin{align}
	& \Big[ -\frac{{\hbar^2}}{2}\sum_{i}{\nabla'}_i^2 + \sum_{i}{\frac{1}{\hbar^{2}}}V_{\rm ext}({\frac{1}{\hbar^{2}}}\br'_i) + \sum_{i<j}\frac{1}{|\br'_i-\br'_j|} \Big]\Psi \nl
	& = E'\Psi. \label{SEhbar}
\end{align}
Here $E'\equiv E / \hbar^2$. \Eq{SEhbar} corresponds to an auxiliary system with one-body potential $V'_{\rm ext}(\br) = \frac{1}{\hbar^{2}}V_{\rm ext}({\frac{1}{\hbar^{2}}}\br)$. Compared with the original system, the external potential is squeezed and deepened; see \Fig{schemScaling} (a) for an illustration.
Through introducing this auxiliary potential $V_{\rm ext}'$, one can bridge a strongly correlated system of normal $\hbar =1$ with a semiclassical system with $\hbar \rightarrow 0$; they are equivalent in the sense that their wave functions (and also electron densities) only differ by a scaling of the coordinates.
As a real-world example, a dissociated normal H$_2$ molecule with internuclear distance $R = 74 \,\AA$ is equivalent to a compact H$_2$ with $R' = 0.74 \,\AA$ and scaled $\hbar = 0.1$, as shown in \Fig{schemScaling} (b).
The scaled system with normal bondlength but deepened potential is as strongly correlated as the dissociated H$_2$ with normal $\hbar=1$.
Therefore, looking into the scaled system from the semiclassical perspective shall give us good insight into the increasingly correlated physics as the molecule approaches its dissociation limit.
Moreover, when evaluating DFAs for the scaled density, we shall invoke the inverse mapping from the scaled system to the normal system with $\hbar = 1$ and plug in the unscaled density, which will be discussed in subsequent subsections.

\begin{figure}[htpb]
	\begin{center}
		\includegraphics[width=\linewidth]{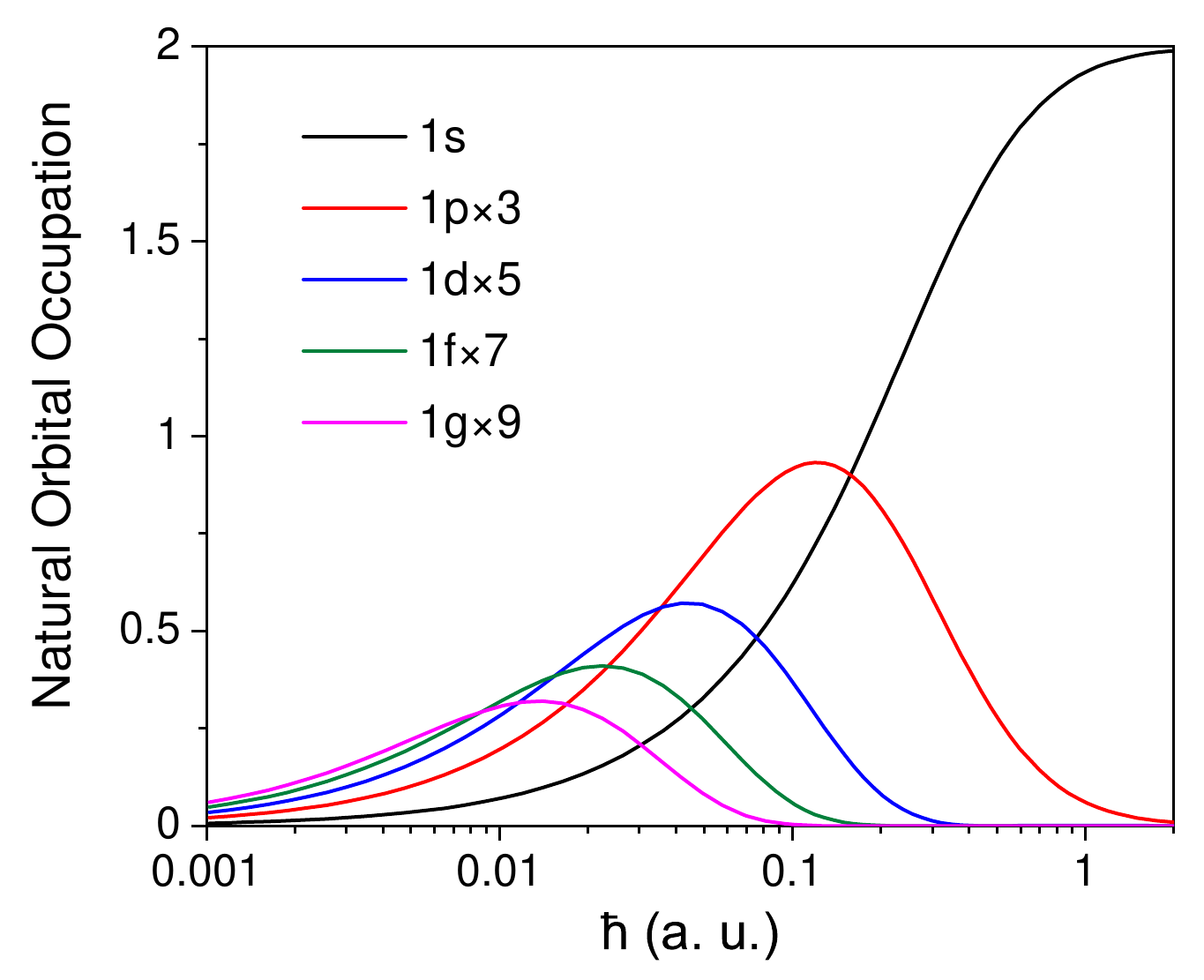}
	\end{center}
	\caption{Evolution of the occupation number distribution over
		selected natural orbitals (multiplied by their respective degeneracy) with $\hbar$ for the ground state of harmonium ($\omega=0.373$). } \label{HookesHeNO}
\end{figure}

For correlations that occur in a compact molecule such as a metal oxide, however, the above scaling argument cannot lead to useful semiclassical picture. For such systems,
a useful way of introducing effective $\hbar$ is to map the system to an exactly solvable model problem with adjustable $\hbar$. In particular, by tuning the $\hbar$ of the model system until one finds the largest resemblance with the real molecular system in the natural occupation distributions, one can define this particular $\hbar$ to be the effective $\hbar$ of the molecule of interest.
In this work, we choose as our model system the harmonium problem (also known as the Hooke's helium), whose Hamiltonian is given by
\begin{align}
	\hat{H} = -\frac{\hbar^2}{2}(\nabla^2_1 + \nabla_2^2) + \frac{1}{2} \omega^2 (r_1^2 + r_2^2) + \frac{1}{r_{12}}.
\end{align}
In our recent work, we have obtained the following compact analytic formula for its ground state:\cite{Yao2446138}
\begin{align}
	\Psi(\br_1,\br_2) = C \Big(1 + \frac{r_{12}}{c}\Big)^\gamma e^{-\frac{\omega(r_1^2 + r_2^2)}{2\hbar}} G\Big(\frac{r_{12}}{r_{12}+c}\Big), \label{WF-harmonium}
\end{align}
Here $C$ is the normalization constant, $\gamma = \frac{E}{\hbar\omega}-3$ with $E$ being the energy, and $c>0$ is a parameter that one can choose in a reasonable range; $G$ is a series function whose expansion coefficients (dependent on $\hbar, \omega$ and $c$) can be analytically obtained. \Eq{WF-harmonium} is valid for any parameter $\hbar$ and $\omega$ and can be evaluated to arbitrary accuracy with a reasonable cost. For some particular parameters, $G$ reduces to a polynomial, as was found in the literature. \cite{Taut933561}
Here of particular interest is the distribution of natural occupation numbers and its evolution with $\hbar$, as it is an indicator of how strongly electrons are correlated. With the analytic expression of \Eq{WF-harmonium}, this can be computed very easily; the result is  shown in \Fig{HookesHeNO}.
When $\hbar$ is large, essentially only the 1s orbital is doubly occupied, which is a weakly correlated regime where the wave function is well described by a single Slater determinant.
When $\hbar$ is small and approaches 0, \Fig{HookesHeNO} shows that all natural orbitals are fractionally occupied. Moreover, each orbital is occupied by a tiny little bit whereas these tiny numbers sum up to an integer 2.
If we translate this picture to the language of CI expansion, it means that the wave function is composed of infinitely many Slater determinants while none of them dominates. This is qualitatively different from the conventional static correlation picture where the wave function is always dominated by a finite number of determinants, such as in a stretched molecule.
As a consequence, the semiclassical limit of harmonium (in fact for other systems as well) is an extremely hard situation for conventional multi-configuration based methods such as MRCI or CASSCF; as shown in  Ref \cite{Yao2446138}, one needs an astronomical number of determinants to obtain a reasonable accuracy.
With our subsequent semiclassical analysis of \Eq{HOA-E}, however, the correct semiclassical limit can be easily recovered.
This suggests that with the very different physical pictures as in \Fig{schemWCvsSC}(a) and (b), it might be beneficial to use distinct starting point to develop theory.

\begin{figure}[htpb]
	\begin{center}
		\includegraphics[width=\linewidth]{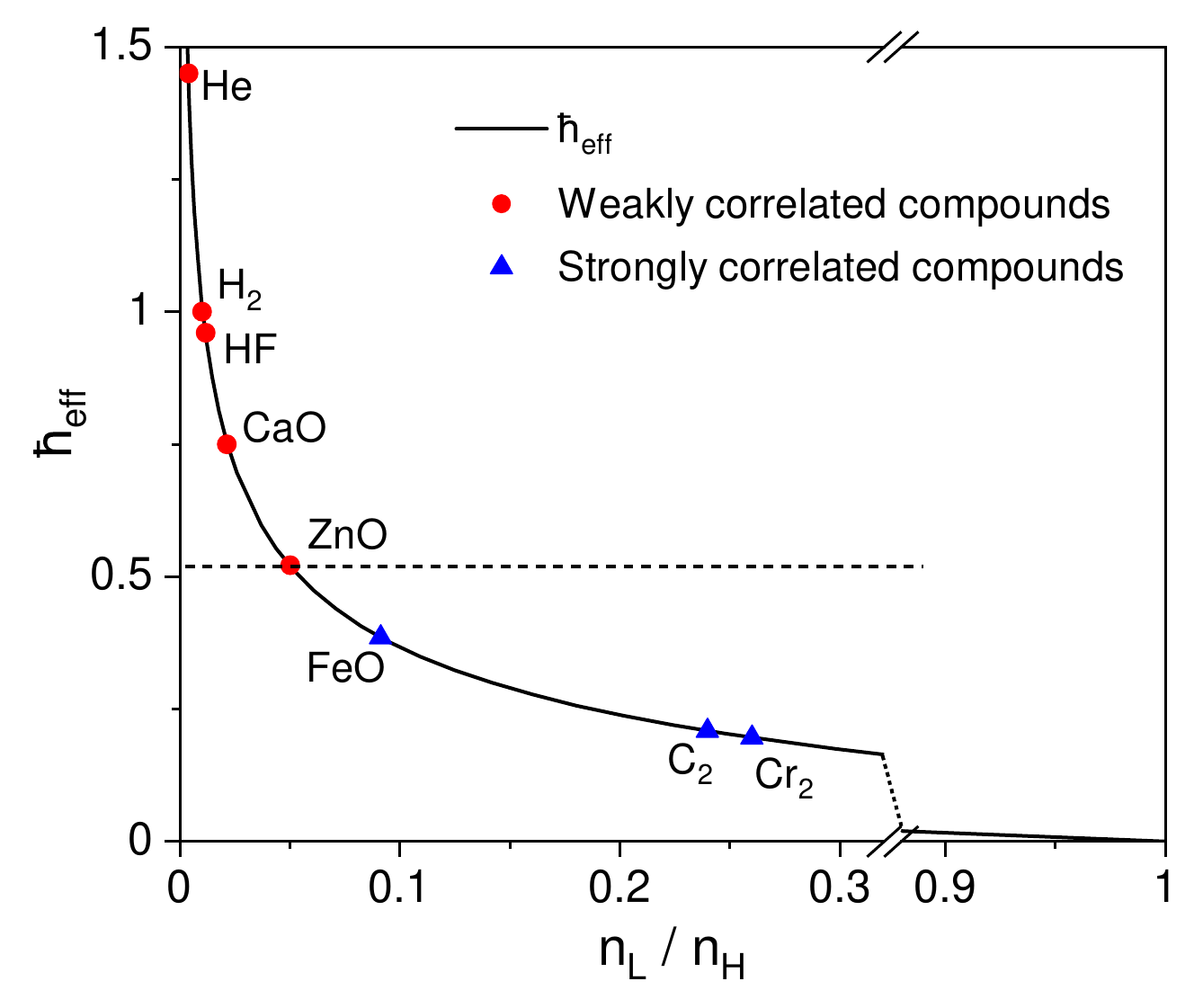}
	\end{center}
	\caption{
		Benchmarking curve showing the mapping relation between $\hbar_{\text{eff}}$ and $\tau = \frac{n_L}{n_H}$, which is obtained from the model harmonium calculation with $\omega = 0.373$. Marked along the curve are some representative atoms and diatomic molecules.
	} \label{nHnLhbar}
\end{figure}

The criteria for defining the largest resemblance of natural occupation distribution is not unique.
Although such a distribution is system dependent, we highlight the following general structures: aligned in decreasing order,
the natural occupations of a general system is given by $1,1,1,\cdots, n_H, n_L, \cdots, 0, 0, \cdots$, where the largest numbers are close to 1 and the smallest numbers close to 0 and all these numbers sum up to the total number of electrons $N$.
In this work, we choose $\tau = \frac{n_L}{n_H}$ as the mapping variable that bridges a real-world system to the model harmonium problem, where
$n_H$ and $n_L$ are the occupation number of the $N$th and $(N+1)$th spin-dependent natural orbitals, respectively.
The corresponding natural orbitals can be viewed as analogues of the highest occupied molecular orbital (HOMO) and lowest unoccupied molecular orbital (LUMO).
Moreover, we choose $\omega = 0.373$ for the benchmarking harmonium such that when $\hbar = 1$ the mapping variable $\tau$ agrees with that of a normal H$_2$ at bondlength $R = 0.74 \AA$.
The resulting benchmarking curve of $\heff$ as a function of $\tau$ is shown in \Fig{nHnLhbar}, where we highlight the small $\tau$ region relevant for chemistry.
For a real-world molecule, by performing a CASSCF calculation we can compute the approximate many-body wave function and obtain its natural occupation numbers and $\tau$. By looking up the value of this particular $\tau$ in \Fig{nHnLhbar}, we can read out its $\hbar_{\rm eff}$.
Some representative atom and diatomic molecules have been marked along the benchmarking curve. For example, by our definition H$_2$ molecule with $\hbar_{\rm eff} = 1$ is a typical weakly-correlated system; He atom with $\hbar_{\rm eff} = 1.45$ has even weaker correlation.
ZnO with $\hbar_{\rm eff} \approx 0.5 $ lies at the boundary line between weakly and strongly correlated systems. With even smaller $\hbar_{\rm eff}$ of about 0.2, C$_2$ and Cr$_2$ are typical strongly correlated systems. These results are in line with our chemical intuition.

\begin{figure}[htpb]
	\begin{center}
		\includegraphics[width=\linewidth]{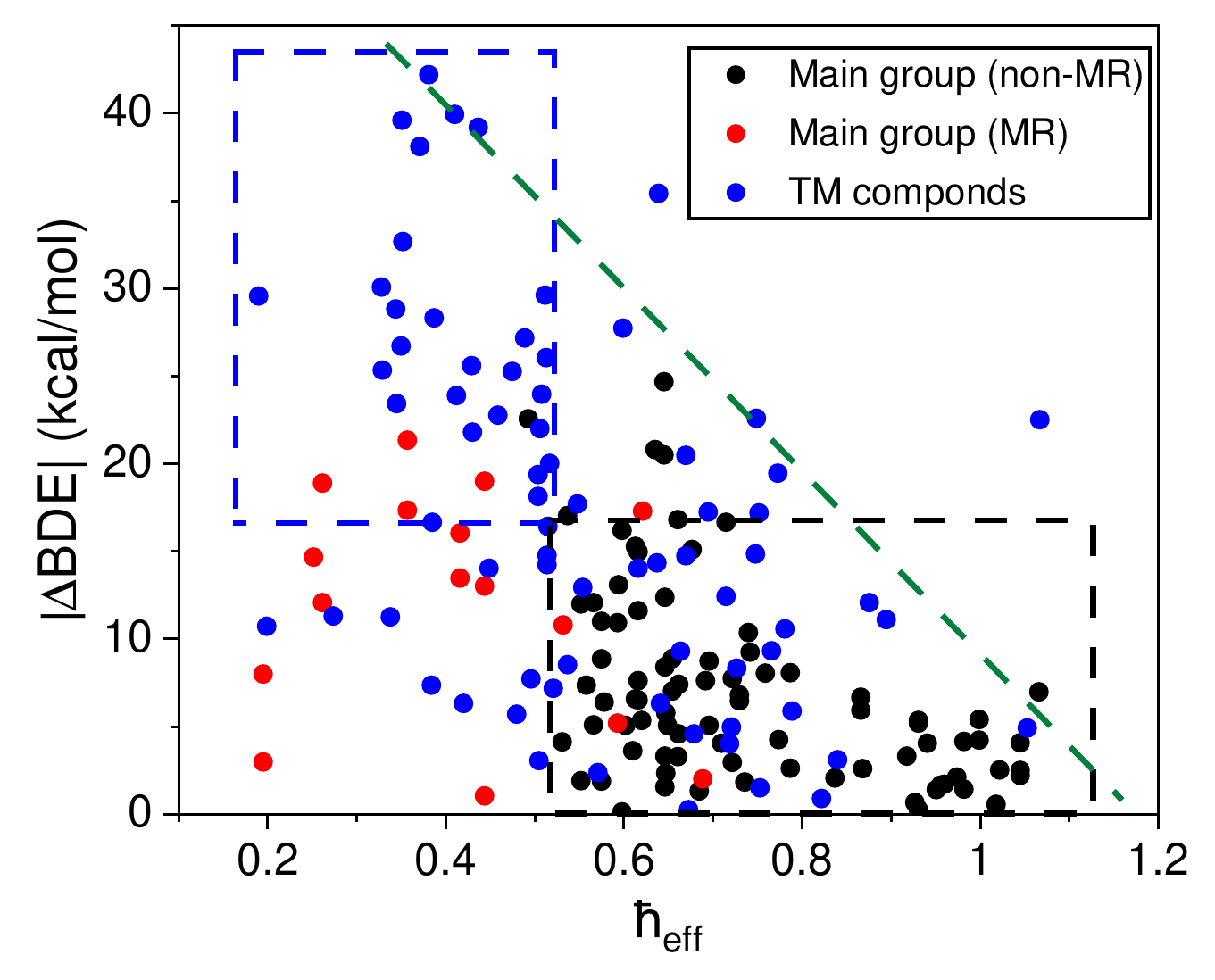}
	\end{center}
	\caption{
		Scatter plot illustrating the correlated relation between $\heff$ and the error of PBE functional on the prediction of bond dissociation energies (BDEs).
		Black and red dots: nonmultireference and multireference subsets, respectively, of the BDE99 dataset \cite{Mardirossian172315,Karton11165} composed of main group elements.
		Blue dots: diatomic molecules containing third-row transition-metals (TMs), including TM hydrides,\cite{Moltved183479} oxides, \cite{Moltved1918432} nitrides and dimers \cite{Furche060021}.
		The dashed line and boxes serve as guides of the eye, highlighting the general trend; see the text for details.
	} \label{dBDEhbar}
\end{figure}

To establish a semi-quantitative relation between the error of DFAs with the magnitude of correlation of small and medium-sized molecules, in \Fig{nHnLhbar} we present a scatter plot of the bond dissociation energy (BDE) error of the Perdew-Burke-Enzerhof (PBE) functional\cite{Perdew963865} against $\heff$ for 172 data points from the well-known datasets documented in the literature.\cite{Mardirossian172315,Karton11165,Moltved183479,Moltved1918432,Furche060021}
The scatter plot for LDA shows similar trend, see the supplemental information (SI) for details. \cite{supp}
As shown in \Fig{dBDEhbar}, $\heff$ of these molecules range from 0.2 to 1.1, while the energy errors range from a few kcal/mol up to 42 kcal/mol.
We highlight that the distribution of these data points are not completely random, instead, most of them scatter in a lower triangular region underneath the green reference line.
This suggests that the general trend is that the error drops as $\heff$ increases.
Moreover, we have drawn two dashed boxes in black and blue, respectively, which contain the majority of the data points.
Specifically, most of the weakly correlated main group compounds (black dots) lie in the black box where $\heff$ is closer to 1 rather than 0, and the PBE errors are relatively small.
These account for the majority (60\%) of the molecules tested, where PBE works well.
By contrast, molecules with a large error (over 16.5 kcal/mol) account for 28\% of the data points, and a large proportion of these molecules (mainly transition-metal compound marked as blue dots), falling into the blue box, are strongly correlated in the sense of having $\heff$ closer to 0 rather than 1, and their energies are all underestimated.
As a side remark, some molecules (red dots) in our examined dataset composed of main group elements have been classified as having multireference character. Here in our plot, they have smaller $\heff$ in the general trend than those that are classified as in the non-multireference subset (black dots).
However, the errors of these red dots are not significantly larger than the black dots. We believe this is because some strong correlation effect has already been built in the PBE correlation functional through satisfying some limiting conditions, such as the low density limit of HEG, which allows it to partially capture the correlation for non-uniform densities. This could also be used to explain the blue dots that fall outside the black and blue box.
Despite these outliers, we emphasize that molecules in the blue box are our targets, for which we attempt to understand their errors through semiclassical analysis of the functional forms of mainstream DFAs.

\subsection{Exact behavior of energy and density at the semiclassical limit}

Let us group all the potential energy in \Eq{SE1} as $V_{\rm tot} = \sum_{i=1}^{N}V_{\rm ext}(\br_i) + \sum_{i<j} \frac{1}{r_{ij}}$.
Assuming for a moment that $V_{\rm ext}(\br)$ is bounded from below such that the many-body potential $V_{\rm tot}$ has a minimum,
which is the case for harmonium.
In the limit $\hbar \rightarrow 0$, solving the quantum mechanical ground state energy of such a system reduces to the minimization of $V_{\rm tot}$, 
and the ground state electron density becomes a sum of $\delta$-functions,
\begin{align}
	\rho(\br) = \sum_{i} \delta(\br - \br_i^0), \label{SC-rho}
\end{align}
where $(\br_1^0, \br_2^0, ... \br_N^0)$ is the minimizer of $V_{\rm tot}$.
In the presence of degeneracy induced by rotation symmetry such as in harmonium, the summation in \Eq{SC-rho} shall be replaced by integration.
The physical picture is shown in \Fig{schemWignerCrystal}(b), where electrons condense at their classical positions.

When $\hbar $ is not zero but sufficiently small, the above picture shall be amended to allow electronic vibrations within a narrow range around their classical positions, analogous to the harmonic oscillator approximation (HOA) of nuclear vibrational problems.
In particular, by expanding $V_{\rm tot}$ in Taylor series around the classical positions up to the second order and then diagonalizing the Hessian matrix, one can obtain the normal modes and the corresponding frequencies $\omega$.
It can be shown that the energy obtained within HOA, call it $E_{\rm gs}^{\text{HOA}}$, captures the leading order of $\hbar$ correction to the total energy, \cite{supp}
\begin{align}
	 E_{\rm gs}^{\text{exact}} = E_{\rm gs}^{\text{HOA}} + \mO(\hbar^2). \label{HOA-E}
\end{align}
The electron density is also broadened from a sum of $\delta$-functions to Gaussian-type distributions around the classical positions.

\subsection{DFA behaviors in semiclassical atomic models}

\begin{figure}[htpb]
	\begin{center}
		\includegraphics[width=\linewidth]{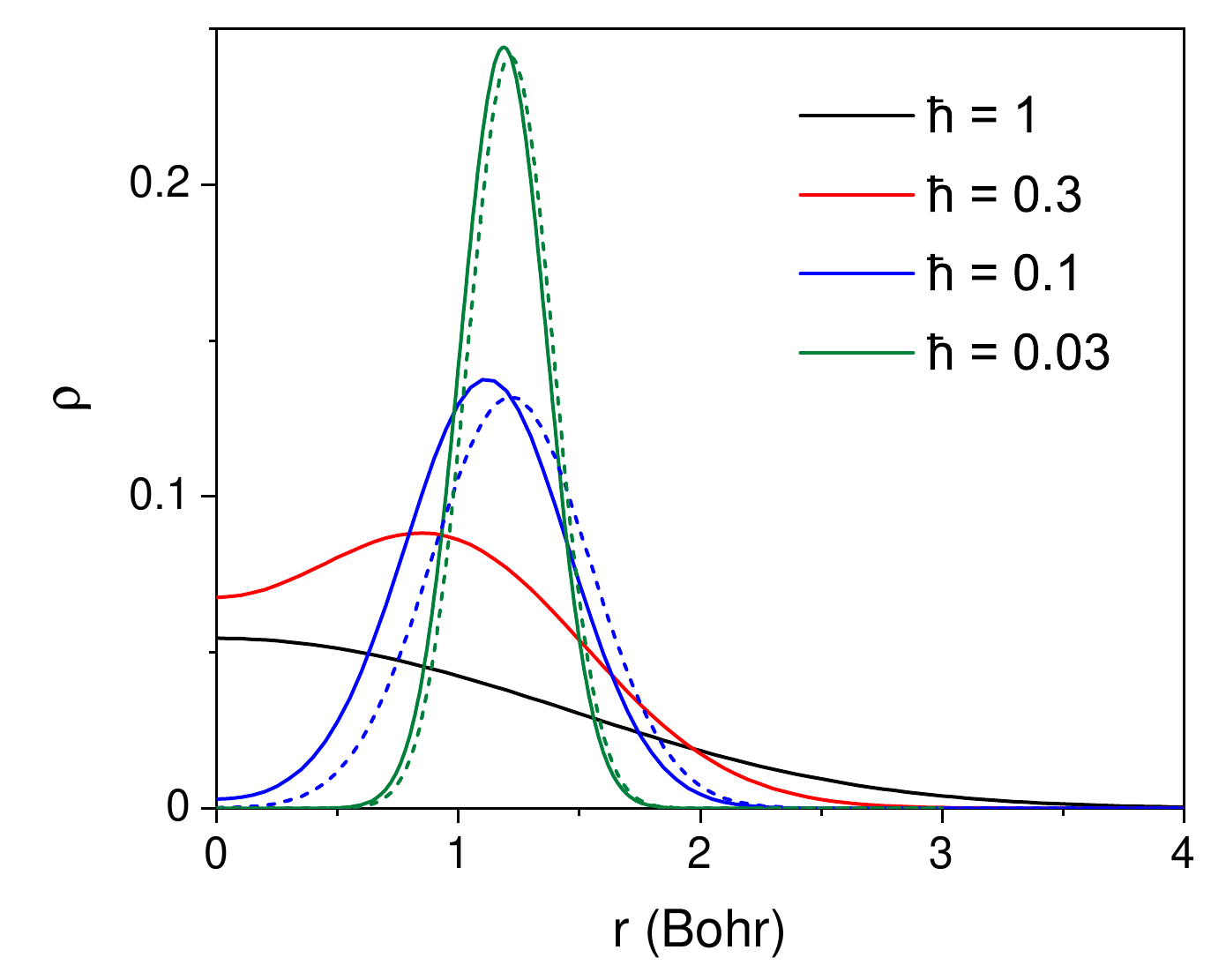}
	\end{center}
	\caption{
		Evolution of the exact ground state density (solid lines) of harmonium with decreasing $\hbar$. The HOA densities (dashed lines) calculated by \Eq{rhoA} are also shown for some small $\hbar$ for comparison.
	} \label{HookesHeRho}
\end{figure}

The harmonium problem provides us an ideal playground for testing the performance of DFAs in the semiclassical limit.
Using the exact wave function as given by \Eq{WF-harmonium}, we can compute the exact density. One can show that the density has spherical symmetry; by integrating out the Euler angles, it can be expressed as the following integration over $r'$:
\begin{align}
	\rho(\br) = & \frac{B e^{-2\alpha r^2}}{r}  \int_0^\infty dr' \nl
	& \times r' (1 + \frac{r'}{c})^{2\gamma} e^{-\alpha r'^2}\sinh(2\alpha r r')G^2\Big(\frac{r'}{r'+c}\Big). \label{rho}
\end{align}
Here $\alpha = \frac{\omega}{\hbar}$ and $B=8C^2 \sqrt{\frac{2\alpha}{\pi}}$. Shown in \Fig{HookesHeRho} is the evolution of the exact density with $\hbar$. For weakly correlated case such as when $\hbar =1$, the density peaks at the origin. As $\hbar$ decreases, the enhanced electron correlation induces a shoulder in the density profile, which gradually develops into a Gaussian-type distribution for diminishing $\hbar$.
This is in line with our HOA analysis. In particular, one can deduce the following Gaussian behavior for small $\hbar$,
\begin{align}
	\rho^{\rm HOA}(\br) \propto \exp\Big[ -\frac{\omega_0}{\hbar}(r - r_0)^2 \Big], \label{rhoA}
\end{align}
where $r_0 = (2\omega)^{-2/3}$ is the semiclassical electron-nuclear distance and $\omega_0=(3-\sqrt{3})\omega$ is a radial harmonic frequency extracted from the normal mode.
Our HOA analysis has been confirmed numerically; see the increasingly better agreement between the dashed and solid lines for diminishing $\hbar$ as shown in \Fig{HookesHeRho}: $\rho^{\rm HOA}$ is an extremely good approximation to the exact density for small $\hbar$; and as $\hbar \rightarrow 0$, \Eq{rhoA} approaches a $\delta$-function.

With the exact density, we plug it into DFAs and study the functional error in the semiclassical limit.
As we have suggested in subsection A, one needs to extend the xc functionals, which were designed for $\hbar=1$, to arbitrary $\hbar$.
This is achieved through the coordinate scaling as illustrated in  \Fig{schemScaling}. For a system with $\hbar \neq 1$ and density $\rho_\hbar$, we can map it back to a normal system with $\hbar = 1$ and density $\rho_1$, with the scaling relation $\rho_1(\br) = \hbar^6 \rho_\hbar(\hbar^2 \br)$. The energy $E_1$ can then be evaluated by plugging $\rho_1$ into the functional form of DFAs. Finally from the energy scaling argument we can obtain $E_\hbar = \frac{E_1}{\hbar^2}$.
After simplifying the resulting energy expression and singling out the xc functional, we find that it satisfies the following scaling relation: \cite{supp}
\begin{align}
	E_{xc,\hbar}[\rho_\hbar(\br)] = \frac{1}{\hbar^2} E_{xc,1}[\rho_1(\br)]. \label{Exchbar}
\end{align}

\begin{figure}[htpb]
	\begin{center}
		\includegraphics[width=\linewidth]{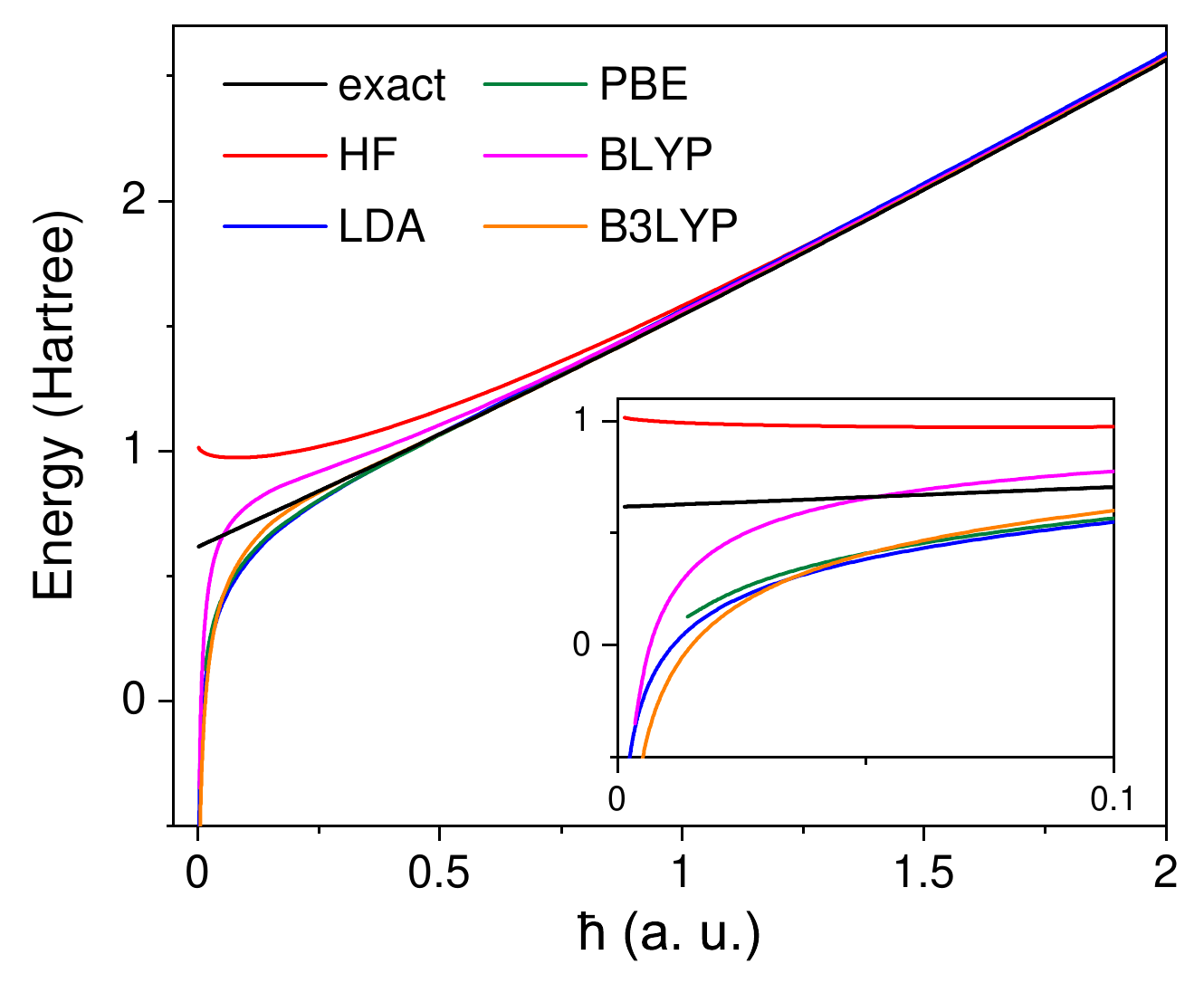}
	\end{center}
	\caption{
		Ground state energy of harmonium (with $\omega = 0.373$) as function of $\hbar$ calculated with some mainstream DFAs, in comparison with the exact one. The inset shows an enlarged plot for the small-$\hbar$ region. PBE results for very small $\hbar$ are not shown due to the limited numerical precision of Libxc.\cite{Lehtola181}
	} \label{HookesHeE}
\end{figure}

The total energy of DFAs as function of $\hbar$
are shown in \Fig{HookesHeE}. Compared with the exact curve, DFAs perform very well for the most part (all lines almost overlap each other), particularly when $\hbar$ is not small.
This is because for weakly correlated systems with 2 electrons, the xc energy is a small term compared to the other three Kohn-Sham energy components that are defined as known and explicit functionals of the density, namely the Kohn-Sham kinetic energy, external energy and the Hartree energy.
Then essentially any mainstream DFA can be a decent approximation to the total energy.
For small $\hbar$, however, divergent behavior starts to emerge. In particular, LDA, GGAs examplified by PBE and the Lee--Yang--Parr (BLYP) functional, and the hybrid B3LYP functional \cite{Becke935648,Lee88785} all diverge to minus infinity in the limit $\hbar \rightarrow 0$. By contrast, the Hartree-Fock (HF) functional converges to a finite value because it is derivable from a variational wave function theory and shall always overestimate the exact total energy, which is finite in the semiclassical limit and given by the minimum of $V_{\rm tot}$.

Using HOA, we can derive the asymptotic behavior of LDA exchange as $\mO(\hbar^{-\frac{1}{6}})$. \cite{supp}
This divergent behavior can be attributed to the local exchange energy that is proportional to $\rho^{4/3}$: as $\rho$ approaches $\delta$-function,
$\rho^\alpha$ integrates to infinity for any $\alpha>1$.
Similar analysis can also be performed for the VWN5 correlation functional, although analytic derivation becomes more difficult. By numerical fitting, we find it also diverges as $\mO(\hbar^{-\frac{1}{6}})$. \cite{supp}
Therefore, for LDA we have
\begin{align}
	E^{\text{LDA}}_{\text{atom}} & = C_1 \hbar^{-1/6} + \mO(1) , \label{asympt-atom}
\end{align}
which violates the following exact constraint:
\begin{align}
	E^{\text{exact}}_{\text{atom}} & = C_0 + \mO(\hbar) .  \label{E_exact}
\end{align}

\begin{figure}[htpb]
	\begin{center}
		\includegraphics[width=\linewidth]{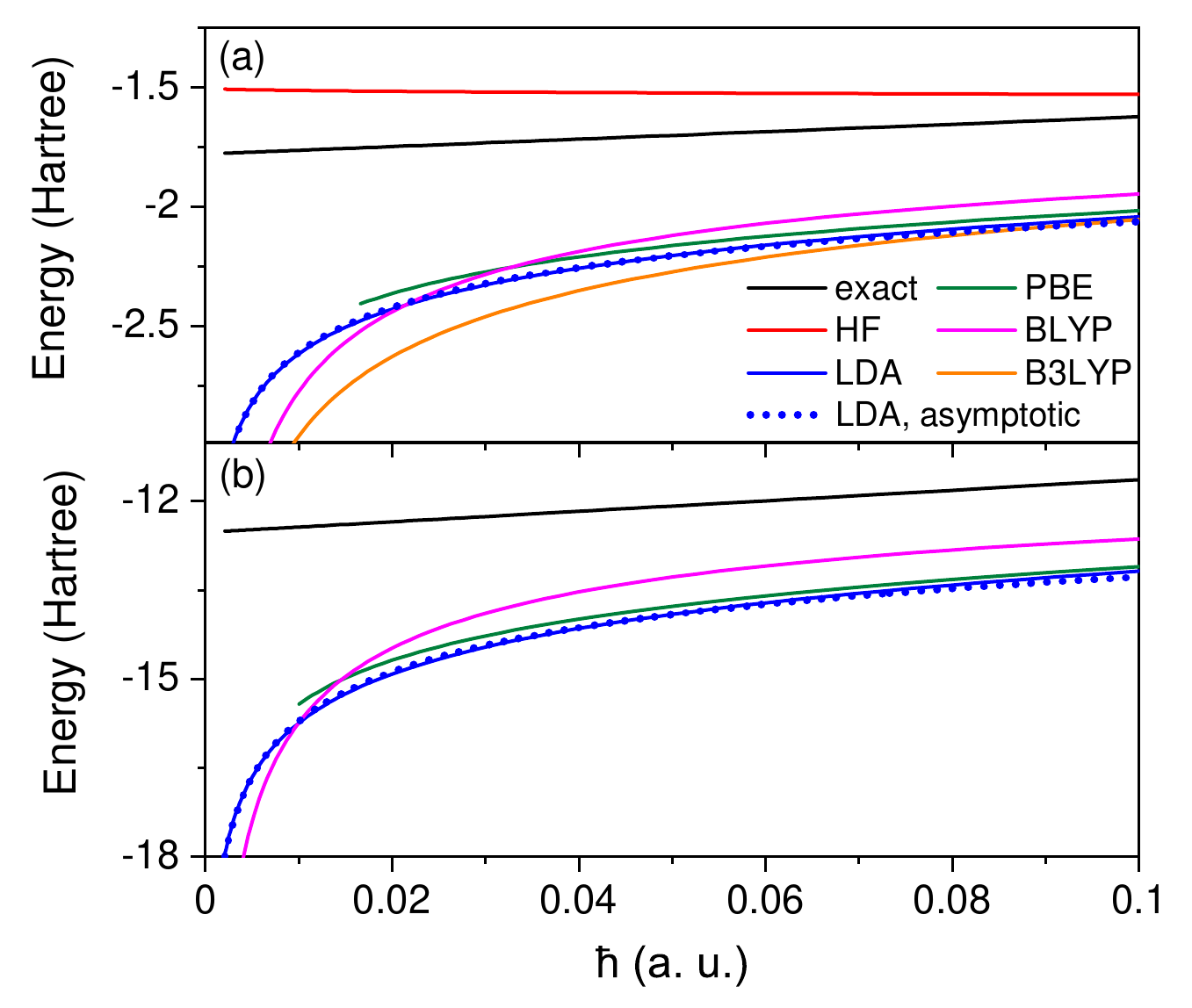}
	\end{center}
	\caption{
		Total energy as function of $\hbar$ in the small-$\hbar$ region for (a) model Mg atom and (b) model As atom under pseudopotential.
		DFAs are compared with the exact curve under HOA.
		The asymptotic expansion of \Eq{asympt-atom} with fitted expansion coefficients up to $\mO(1)$ is shown in blue dotted line for comparison.
		In (b), the Hartree--Fock and B3LYP results are not shown; see our remarks in the text.
	} \label{BLPSatoms}
\end{figure}

For real-world atoms with Coulombic nuclear attraction potentials, our semiclassical analysis cannot be directly applied to \Eq{SE1} because  $V_{\rm ext}$ and also $V_{\rm tot}$ are not bounded from below. Nevertheless, let us remind ourselves that the most essential part of correlation occurs among valence electrons; core electrons such as those occupying 1s orbitals are rather inert and only weakly correlated with other electrons.
The idea of separating the core from the valence electrons has lead to various approximation schemes, such as the pseudopotential method\cite{Schwerdtfeger113143}, the complete active space self-consistent field (CASSCF) method,\cite{Werner802342,Werner815794,Werner855053} and the recently proposed electronic exact factorization in the Fock space\cite{Requist21116401}.
Due to the Pauli exclusion from the core electrons, valence electrons are unlikely to appear in the vicinity of the nuclei; the resulting effective potential (or pseudopotential) replaces the bare Coulomb divergence by a short-range repulsion, giving rise to a potential minimum.

Here we use the bulk-derived local pseudopotential (BLPS)\cite{Huang087109} developed by Carter et al as $V_{\rm ext}$ and perform semiclassical analysis for the valence electrons of magnesium (2 electrons) and arsenic atom (5 electrons), respectively.
Although the exact density can no longer be obtained easily for arbitrary $\hbar$, we can use the density from HOA analysis to generate meaningful results when $\hbar$ is small. 
In particular, when $\hbar \rightarrow 0$, the density is distributed on a single shell for model Mg atom, similar to the harmonium problem, but spread on two spherical shells for model As atom [see \Fig{schemWignerCrystal} (b)]. With a finite $\hbar$, the density broadens into Gaussian-type functions analogous to \Fig{HookesHeRho}, and well approximated by HOA.
By plugging $\rho^{\rm HOA}$ into the functional form of DFAs, 
we can show that this approximation introduces small error that is of higher-order to $\mO(1)$; see \Fig{BLPSatoms}. This is only slightly larger than the HOA error in the exact curve, which has additional $\mO(\hbar)$ term included through harmonic analysis according to \Eq{E_exact}. Comparing with the exact finite behavior in the semiclassical limit, again we can come to the conclusion that DFAs
all diverge as $\mO(\hbar^{-\frac{1}{6}})$, violating the exact constraint similar to the harmonium problem.
As a side remark, we note that in
the evaluation of DFAs, the Kohn--Sham kinetic energy is neglected because it is $\mO(\hbar)$.
Besides, the Hartree--Fock exchange energy is an explicit functional of the Kohn--Sham orbitals rather than the density.
From the density alone for more than 2 electrons, one needs to invoke the optimized effective potential (OEP)\cite{Talman7636,Sharp53317} techniques to obtain HF or B3LYP results, which are not the focus of the present paper and thus not shown.
For spin unpolarized 2-electron systems, the doubly occupied Kohn--Sham orbital is the square root of the density so that HF exchange can be computed straightforwardly.
Details can be found in the SI. \cite{supp}

\begin{figure}[htpb]
	\begin{center}
		\includegraphics[width=\linewidth]{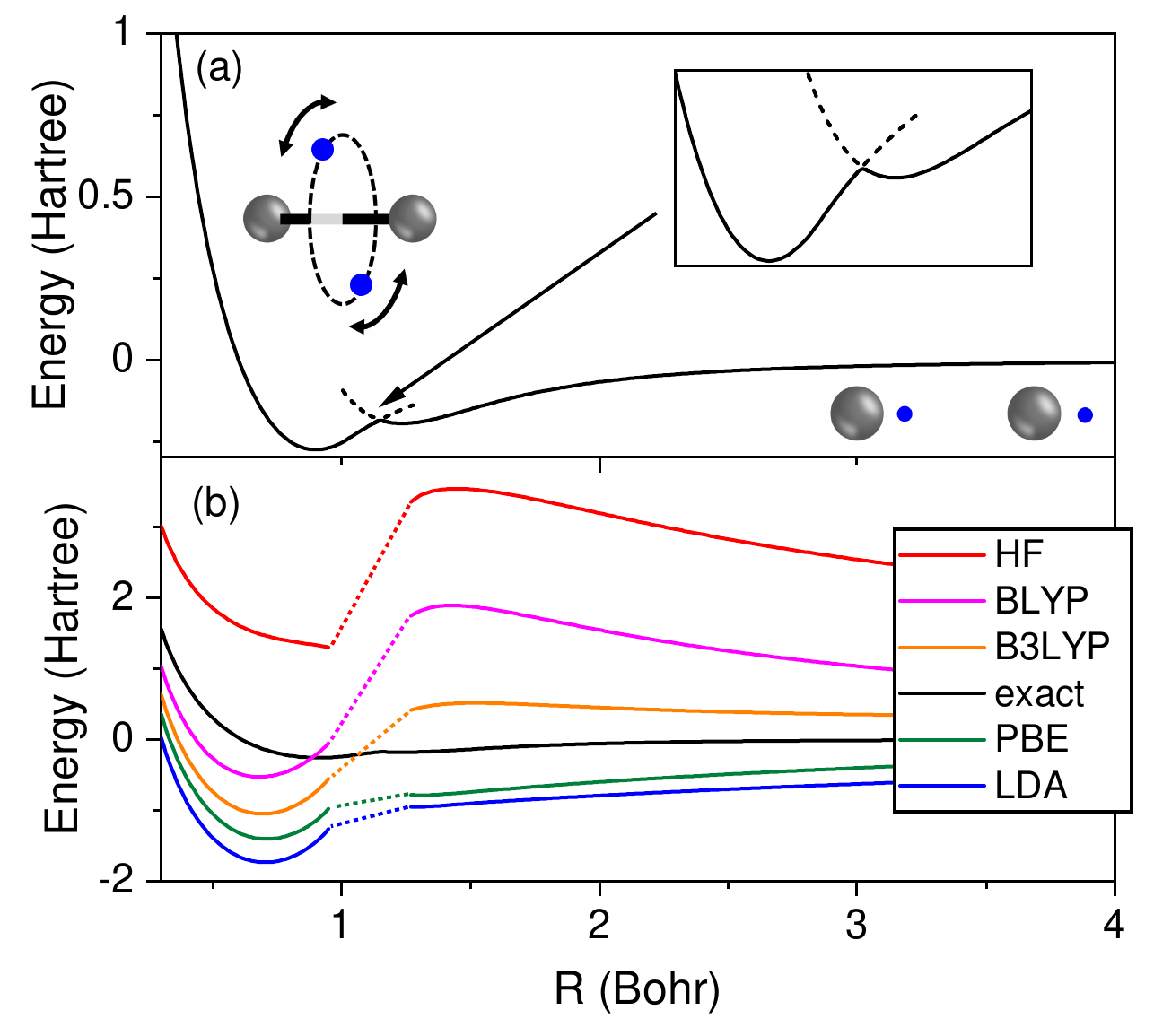}
	\end{center}
	\caption{
		(a) Dissociation energy curve along with the semiclassical pictures showing bonding and bond-breaking regimes of a model 2-electron diatomic molecule as $\hbar \rightarrow 0$.
		The adiabatic energy curve is shown in solid line whereas the diabatic curves are displayed in dashed lines near the crossing point.
		(b) Ground state energy of the same model with $\hbar = 0.01$. Here DFAs under HOA are compared with the exact curve that is accurate up to $\mO(\hbar)$. The DFA results in the dotted region are not shown; see remarks in the text.
	} \label{H2model}
\end{figure}

\subsection{DFA behaviors in semiclassical molecular models}

Our semiclassical analysis can also be performed for molecules. As an example, let us consider a single-bond stretching problem of a diatomic molecule. Again, we focus on the two bonding electrons, which are modeled through
our self-designed effective electron-nuclear potential. \cite{supp}
In contrast to the atomic model, here the change of $V_{\rm tot}(\br_1, \br_2, R)$ with the internuclear distance $R$ leads to interesting physical picture of bond-forming and bond-breaking in the semiclassical limit.
In particular, for small $R$, minimization of $V_{\rm tot}$ with respect to the electronic variables gives a bonding picture where the two electrons localize on the opposite sides of a ring that bisects the two nuclei, reminiscent of the shared electron pair in our routinely used Lewis structure.
For large $R$, bonding picture is energetically unfavorable; instead, the minimum energy is achieved by localizing the electrons in the vicinity of the nuclei, giving a typical bond-breaking picture.
The resulting semiclassical energy as a function of $R$ along with the schematic pictures are shown in \Fig{H2model} (a). As one can expect, there is a critical $R$ at the transition of the two pictures, corresponding to a crossing point of the diabatic curves and a cusp in the adiabatic curve.
For finite $\hbar$, this cusp is rounded out; see for example the black curve in \Fig{H2model} (b).
We highlight that such a shift of picture is rather insensitive to the choice of effective electron-nuclear potential, although it does affect the fine details of transition.
As an additional remark,
the emergence of the critical point in the semiclassical limit reminds us of the Coulson–-Fischer point \cite{Coulson49386} well-known in the dissociation of H$_2$ for normal $\hbar$, where the unrestricted Hartree-Fock calculation switches between two types of solutions. We suspect that there is an internal connection and that the ``phase transition'' type of behavior might be widespread in the dissociation of molecules and may trigger new understanding of chemical bonds.
For example, the peculiar behavior of the dissociation energy curve of chromium dimer\cite{Larsson2215932} might be due to smoothed-out ``phase transitions'' of multiple bond-breaking stages. We leave this for future investigations.

\begin{table}[htpb]
	\caption{
		Asymptotic leading order (with its sign specified if possible) of LDA energy components in the limit $\hbar\to 0$ for model atoms and diatomic molecules. 
	}
	\begin{ruledtabular}
		\begin{tabular}{ccc}
			& Atom & Diatomic molecule \\
			\colrule
			$V_{\rm ext}$ & $\mO(1)$ & $\mO(1)$  \\
			$T_s$ & $+\mO(\hbar)$ & $+\mO(\hbar)$  \\
			$J$ & $+\mO(1)$ & $+\mO(|\ln\hbar|)$   \\
			$E_{xc}^{\rm LDA}$ & $-\mO(\hbar^{-\frac{1}{6}})$ & $-\mO(\hbar^{-\frac{1}{3}})$ \\
			$E_{\rm tot}^{\rm LDA}$ & $-\mO(\hbar^{-\frac{1}{6}})$ & $-\mO(\hbar^{-\frac{1}{3}})$ \\
			$E_{\rm tot}^{\rm exact}$ & $\mO(1)$ & $\mO(1)$  \\
		\end{tabular}
	\end{ruledtabular}
	\label{leading}
\end{table}

Inspired by the exact semiclassical solution, we next perform semiclassical analysis for DFAs with $\hbar=0.01$; see \Fig{H2model} (b).
Depending on the semiclassical picture of bonding or bond-breaking, the HOA density can be qualitatively different. Thus we discuss these two regimes separately.
For small $R$, DFAs typically underestimate the total energy.
In analogy to the atomic model, one can rationalize this deviation by deriving the asymptotic expansion of the LDA energy with $\hbar$ for bonded diatomic molecules as \cite{supp}
\begin{align}
	E^{\text{LDA}}_{\text{mol}} & = C_1 \hbar^{-1/3} + \mO(|\ln\hbar|). \label{asympt-mol}
\end{align}
Here the leading terms are due to the xc and Hartree term, respectively.
Further details of the asymptotic leading order of each component of the LDA functional is tabulated in \Tab{leading}, where we have also compared between atomic and diatomic molecular models.
As in the case of atoms, here LDA violates the exact constraint of finite total energy in approaching the semiclassical limit, and other DFAs suffer from similar failures.
As $R$ is stretched towards the diabatic crossing point, the anharmonic terms of $V_{\rm tot}$ become increasingly important such that our HOA density becomes a bad approximation to the exact density.
In such cases, we have not found simple alternative ways of approximating the exact density to plug into the functional forms of DFAs.
Therefore, DFA results in this part of the dissociation curve are not shown, instead, we replace them with dotted lines.
Beyond this region, HOA revives as a valid approximation, but gives distinct density as suggested in \Fig{H2model} (a).
For such density, DFAs exhibit diversified behaviors, which is not our focus. In the dissociation limit, they all overestimate the total energy (not shown), in line with the fractional-spin analysis. \cite{Cohen08792}

Comparing \Eq{asympt-atom} and \Eq{asympt-mol}, atoms and molecules have distinct semiclassical divergent behaviors for LDA.
In particular, the leading term of $-\mO(\hbar^{-\frac{1}{3}})$ for diatomic molecules diverges faster than the atomic behavior of $-\mO(\hbar^{-\frac{1}{6}})$. This is because the molecular density is more concentrated- two electrons spread their density as delta function on a ring rather than on two spherical shells around the two nuclei, which accelerates the divergence of the xc integral.
As a consequence, one can immediately deduce that LDA always overbinds diatomic molecules when $\hbar$ is sufficiently small.
For $\hbar$ that is not small, unsurprisingly it is well-known that LDA tends to overbind real-world molecules with $\hbar = 1$.\cite{Kohn9612974, Mardirossian172315,Karton11165,Moltved183479}
Previously this has been understood from the perspective of delocalization error. \cite{Cohen08792}
Here we provide a novel perspective which ascribes the overbinding to the violation of an exact constraint at the semiclassical limit.
A further fact that supports our argument is the systematic and severe overbinding of LDA among strongly correlated transition metal diatomic molecules, for which an LDA+$U$ correction is often needed.\cite{Anisimov91943,Liechtenstein95R5467,Kulik100021,Kulik150021} The empirical $U$ correction is positive in most cases, and the more strongly correlated systems demand a larger $U$ correction. This can be well explained using our semiclassical argument: the stronger the correlation, the smaller the $\hbar_{\rm eff}$, which pushes the system closer to the semiclassical limit where the violation of the exact constraint has the largest consequence. Our conclusions derived from LDA also apply to GGAs since they have rather similar behaviors.

\section{Conclusion}

In this work, we introduce semiclassical limit to electronic systems,
where the reduced Planck constant $\hbar$ is treated as an adjustable parameter and tends to 0.
Compared with the conventional understanding of strong correlation through multi-configurational picture, our semiclassical perspective along with the concept of $\heff$ provides a novel insightful picture, which could offer a better starting point for method development.
Besides the conceptual conciseness, our semiclassical limit is particularly pertinent to describing the type of strong correlation that involves combinatorial number of ``democratic'' determinants while no one takes the lead.
Such a scenario could arise from the partially-filled degenerate $d$ orbitals of transition metals and near-degenerate orbitals of nearby atoms in their compounds, and  
is an extremely hard situation from the multi-configurational perspective. It is perhaps even harder for perturbation theory if one starts with a mean field. The conventional approaches, which are suitable for treating one or a handful of determinants, are simply too far from capturing the essence of this type of strong correlation physics.
For such problems, instead of using orbitals describing individual motion, we propose to switch to the collective motion of strongly correlated electrons through HOA in the semiclassical treatment, which is a many-body description by design. 

In contrast to another ``semiclassical'' regime discussed by Burke et al., \cite{Elliott08256406,Okun23179,Burke200021,Burke2098,Cangi10235128,Ribeiro180021}
which scales the electron number to infinity upon taking $\hbar \rightarrow 0$,
our semiclassical limit has totally opposite physics of strong correlation rather than weak correlation.
In similar spirit to the Wigner crystal for periodic solids, our semiclassical limit
could be regarded as its extension to finite systems,
where electrons condense at classical positions defined by the minimum of total potential energy.
Instead of explaining why DFAs work well in the weakly correlated limit as examplified by a neutral atom with large nuclear charge,\cite{Fournais18105,Okun23179,Schwinger801827,Constantin11186406,Fournais18105}
here with our semiclassical regime we aim to find the important missing constraint of DFAs responsible for their qualitative failures in strongly correlated molecules.
In particular, DFAs significantly underestimate total energy in the semiclassical limit. For example, the exchange-correlation energy of LDA erroneously diverges to minus infinity as $\mO(\hbar^{-1/6})$ for model atoms and $\mO(\hbar^{-1/3})$ for model diatomic molecules.
This can explain the overbinding of LDA and GGAs for strongly correlated transition metal diatomic molecules since we have estimated that the effective $\hbar$ for the valence electrons of these systems can be quite small.

The concept of $\heff$ proposed in this work comes from the physical consideration that core electrons are well-separated energetically from other electrons and contribute little to the valence correlation that is relevant to chemistry. Therefore, the effective $\hbar$ for valence electrons does not have to agree with the reduced Planck constant for all-electrons. Here we aim to introduce $\heff$ as a quantitative indicator to characterize the extent of correlation of a particular system. This allows us to compare correlations among different systems rather than vaguely referring all of them to as strongly correlated systems. In this sense, $\heff$ can be a useful concept for describing a system, just like other physically sound but empirical chemical concepts such as bond order or hardness of Lewis acids, which can be appealing to chemists although not uniquely defined.
In this work, we define $\heff$ by mapping the natural occupation numbers of real-world molecules to model harmonium problem via the relation $\frac{n_L}{n_H} \rightarrow \hbar $. This particular choice gives reasonable $\heff$ for the molecules tested. Yet, chances are that one might find other mappings that yield better agreement with our chemical intuition.

As a final remark, we expect the exact semiclassical constraint for $E_{xc}$ proposed in this work can give us important hint for improving the energy prediction of transition metal compounds. For these challenging systems, the known exact conditions for strong correlation, such as the well-established flat plane condition for fractional charges and spins,\cite{Mori-Sanchez09066403} have not found an easy path to be applicable.
So far the successful application of these conditions is mainly to alleviate molecular dissociation problems where DFAs can be improved by restoring the underlying condition for local subsystems using the fractional-spin local orbital scaling correction (FSLOSC) functional \cite{Su189678}.
However, the picture of local fractional spins, derived from isolated subsystems perfect valid in the dissociation limit, becomes ambiguous in unstretched molecules where subsystems come together and form chemical bonds.
In such cases, switching to our semiclassical perspective may inspire new functional development. 
Along this line of thinking, a critical task is to extract the information of $\heff$ from DFAs. If $\heff$ is not small, then little correction is needed. Otherwise, a positive correction should be imposed to the xc energy to counterbalance its underestimated behavior.
We leave this for future investigations.

%
%


~\\

{\bf Supporting Information}
See the supplemantal information at xxx for some details.

{\bf Acknowledgment}

The authors appreciate funding support from the National Key Research and Development Program of China (2023YFA1507000) and the National Natural Science Foundation of China (8200906190).

{\bf Conflict of Interest}

We have no conflicts of interest to disclose.

{\bf Data Availability}

The data that support the findings of this study are available from the corresponding authors upon reasonable request.

{\bf References}
\bibliographystyle{achemso}
\makeatletter
\renewcommand\@biblabel[1]{(#1)}
\makeatother
\bibliography{Ref}

\end{document}


\draft

Supporting Information of
%
\vspace{0.1cm}

\begin{center}
{\large \textbf{Exact Constraint of Density Functional Approximations \\ at the Semiclassical Limit}}
\vspace{0.2cm}

Yunzhi Li and Chen Li

{\small
\emph{Beijing National Laboratory for Molecular Sciences, College of Chemistry and Molecular Engineering, Peking University,
Beijing 100871, China
\\}
}

\vspace{0.5cm}

(Dated: \today)

\end{center}

\linespread{0.90}

\tableofcontents



\renewcommand{\theequation}{S\arabic{equation}}
\renewcommand{\thetable}{S\arabic{table}}
\renewcommand{\thefigure}{S\arabic{figure}}

\section{Some details of the semiclassical limit for many-electron systems}

Here we show that in the semiclassical regime ($\hbar$ is small), the ground state energy can be well described by harmonic oscillator approximation (HOA), with the error on the order of $\hbar^2$.
We begin with the following Schr\"odinger equation,
\begin{align}
	\hat H \Psi \equiv \Big[-\frac{\hbar^2}{2}\sum_{i=1}^N \nabla_i^2 + \sum_{i=1}^N V_{\rm ext}(\br_i) + \sum_{i<j}\frac{1}{r_{ij}}\Big]\Psi = E\Psi.
\end{align}
Here $V_{\rm tot} \equiv \sum_{i=1}^{N}V_{\rm ext}(\br_i) + \sum_{i<j} \frac{1}{r_{ij}}$ denotes the total potential energy.
For the ease of discussion, let us first consider the spinless problem. Later, we will come back to the problem with spin and show that the spin degrees of freedom only affect higher-order terms in the semiclassical expansion.

Assuming that minimization of $V_{\rm tot}$ is achieved at $\bm u=(\br_1^0, \br_2^0, ... \br_N^0)$, yielding minimum potential energy $V_0$. Here $\br^0_i \neq \br^0_j$ for any $i\neq j$, otherwise $V_0$ cannot be finite.
Because $V_{\rm tot}$ has permutation symmetry, its minimizer is not unique. In particular, any permutation of $\bm u$ (with $N!$ possibilities), denoted as $\bm u_p = (\br_{p_1}^0, \br_{p_2}^0, ... \br_{p_N}^0)$, is also a minimum.
We then perform Taylor expansion at one of these minima, similar to the many-body vibrational problem. Here $V_{\rm tot}$ is analogous to the many-body potential energy surface that depends on $3N$ electronic cartesian coordinates. By harmonic analysis, i.e., diagonalizing the $3N\times 3N$ Hessian, one can obtain the harmonic oscillator frequencies $\omega_k$ and the normal coordinates $q_k$; each $q_k$ is a collective coordinate involving all $3N$ cartesian coordinates. The Schr\"odinger equation then becomes
\begin{align}
	\Big[ -\frac{\hbar^2}{2}\sum_{k=1}^{3N} \frac{\p^2}{\p q_k^2} + \frac{1}{2} \sum_{k=1}^{3N} \omega_k^2 q_k^2 + V_{\rm anh}(q_1, q_2, \cdots, q_{3N}) \Big]\Psi = (E-V_0)\Psi, \label{SEq}
\end{align} 
where $V_{\rm anh}$ is the anharmonic contribution of $V_{\rm tot}$.
In \Eq{SEq}, the kinetic energy operator is a small term in the semiclassical limit, which in principle can be treated perturbatively. However, directly performing perturbation theory with $\hbar$ leads to problematic series.
To avoid this difficulty, we transform \Eq{SEq} into a simpler problem through scaling the collective coordinates by $t_k = \frac{q_k}{\sqrt{\hbar}}$, which gives
\begin{align}
	\Big[ -\frac{1}{2}\sum_{k=1}^{3N} \frac{\p^2}{\p s_k^2} + \frac{1}{2} \sum_{k=1}^{3N} \omega_k^2 t_k^2 + \sqrt \hbar \tilde V_{\rm anh,3} + \hbar \tilde V_{\rm anh,4} + \cdots \Big]\Psi = E'\Psi. \label{SEs}
\end{align}
Here $E' = (E - V_0)/\hbar$; and we have expanded $V_{\rm anh}$ in multi-dimensional Taylor series of $t_k$'s, with the leading order terms being cubic terms, collected as $\sqrt\hbar \tilde V_{\rm anh,3}$, and the next leading terms are quartic terms, collected as $\hbar \tilde V_{\rm anh,4}$, and so on. Neglecting higher orders of $\sqrt \hbar$, we then perform perturbation expansion in $\sqrt \hbar$ for the ground state energy and wave function. 
The zeroth-order wave function (also known as the HOA wave function)
is a product of $3N$ harmonic oscillator functions, given by
\begin{align}
	\Psi_0 = \prod_{k=1}^{3N} \Big(\frac{\omega_k}{\pi\hbar}\Big)^{\frac{1}{4}} \exp\Big(-\frac{1}{2}\omega_k t_k^2\Big), \label{Psi0}
\end{align} 
with the zeroth-order energy
\begin{align}
	E'_0 = \frac{1}{2}\sum_{k} \omega_k.
\end{align}
The 1st-order perturbation correction, given by $\langle \Psi_0 | \sqrt \hbar\tilde V_{\rm anh,3}|\Psi_0\rangle$, shall vanish. This is because $\Psi_0$ has even parity while $\tilde V_{\rm anh,3}$ has odd parity. Therefore, 
the leading correction comes from 2nd-order perturbation so that
\begin{align}
	E' = \frac{1}{2}\sum_{k} \omega_k + \mO(\hbar). \label{Ep}
\end{align}
It follows that 
\begin{align}
	E = V_0 + \frac{\hbar}{2}\sum_{k} \omega_k + \mO(\hbar^2) = E^{\rm  HOA } + \mO(\hbar^2). \label{Ehbar}
\end{align}
Thus, the energy obtained within HOA captures the leading order of $\hbar$ correction to the total energy.

Here as a side remark, because we start with a particular minimizer $\bm u$ of the total potential to perform our HOA analysis and $\bm u$ is not invariant to permutations, our harmonic Hamiltonian no longer has permutation symmetry. This induces a symmetry-breaking HOA wave function $\Psi_0$. If one starts with a different minimizer $\bm u_p$ to perform HOA analysis, one should end up with a different HOA wave function $\Psi_p$. They are related by a simple permutation, i.e., $\Psi_p(\br_1, \br_2, \cdots, \br_N) = \Psi_0(\br_{p_1}, \br_{p_2}, \cdots, \br_{p_N})$. 

Next, let us consider the spin degrees of freedom and construct the full space--spin wave function.
Under HOA, we can construct the antisymmetric full wavefunction through linear combinations of $\Psi_p$,
\begin{align}
	\Phi^{\rm HOA}(\br_1,\br_2,...,\br_N;s_1,s_2,...,s_N) 
	= \frac{1}{\sqrt{N!}} \sum_{p} (-1)^{\tau(p)} \Psi_0(\br_{p_1},\br_{p_2},...,\br_{p_N}) \sigma(s_{p_1},s_{p_2},...,s_{p_N}). \label{sclPsi}
\end{align}
Here $\sigma$ is an arbitrary normalized $N$-electron spin wave function with $s_i$ being the spin of the $i$th electron; $\tau$ is the number of transposition of permutation $p$.
The HOA energy can be computed by taking the following
the expectation value,
\begin{align}
	E^{\rm HOA} = \frac{\la\Phi^{\rm HOA}|\hat{H}|\Phi^{\rm HOA}\ra }{\la\Phi^{\rm HOA}|\Phi^{\rm HOA}\ra }
	= & V_0 + \frac{\hbar}{2}\sum_{k} \omega_k + \mO(\hbar^2) . \label{HOA-anti}
\end{align}
In the derivation of \Eq{HOA-anti}, we have neglected $\la \Psi_p | \Psi_{p'} \ra$ and $\la \Psi_p |\hat {H}| \Psi_{p'} \ra$ for $p\neq p'$ since
$\Psi_0$ approaches a product of $\delta$-functions at distinct locations when $\hbar$ is sufficiently small. \Eq{HOA-anti} is essentially the same as our spinless HOA result. This means that spin only affects higher-order terms of $\hbar$ in the semiclassical limit, which is reasonable because particles behave classically and quantum effects such as spin splitting become negligible. This is also the case for nuclei, where different nuclear spin states differ tiny little bit in energy, detectable only by nuclear magnetic resonance (NMR).

\section{Supplementary results of $\hbar_{\text{eff}}$ calculations}

\begin{figure}[htpb]
	\begin{center}
		\includegraphics[width=0.8\linewidth]{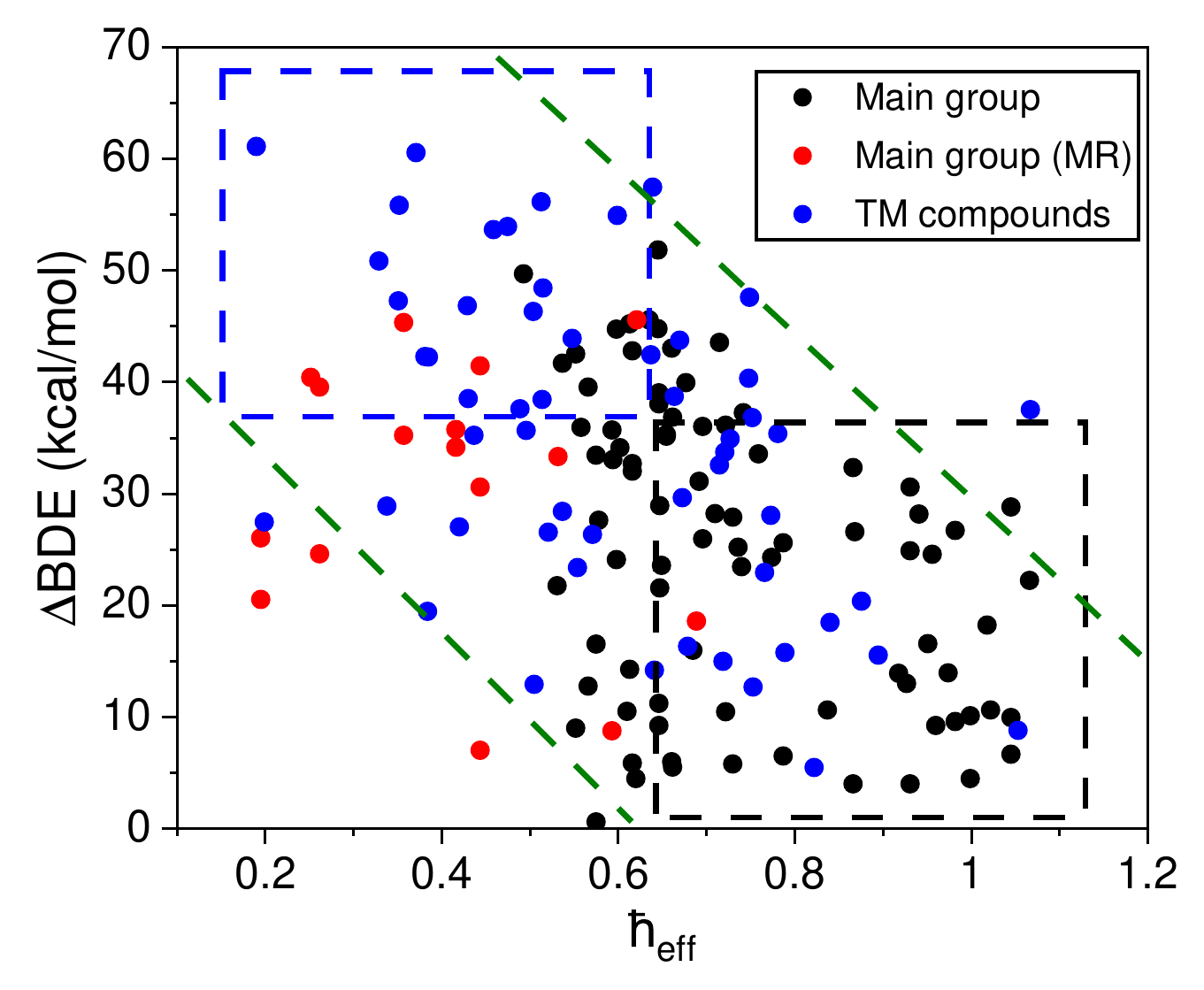}
	\end{center}
	\caption{
		Scatter plot illustrating the correlated relation between $\heff$ and the error of LDA functional in the prediction of bond dissociation energies (BDEs). 
		Black and red dots: nonmultireference and multireference subsets, respectively, of the BDE99 dataset \cite{Mardirossian172315,Karton11165} composed of main group elements.
		Blue dots: diatomic molecules containing third-row transition-metals (TMs), including TM oxides \cite{Moltved1918432}, hydrides \cite{Moltved183479}, nitrides and dimers \cite{Furche060021}. 
		The dashed lines and boxes serve as guides of the eye, highlighting the general trend; see the text for details. 
	} \label{dBDEhbarLDA}
\end{figure}

In order to evaluate $\heff$ for a molecule, we need to calculate $\tau = \frac{n_L}{n_H}$ from the natural occupation numbers. And these numbers are obtained through complete active space self-consistent field (CASSCF) calculations.
Although CASSCF results are known to depend on the choice of active space and basis set, in our practice we find that given a reasonable choice based on our chemical intuition the $N$th and $(N+1)$th natural occupation number ($n_H$ and $n_L$) are not very sensitive to the change of active space or basis set, whereas such changes do have a great influence on those very small natural occupation numbers. 
Therefore, we believe CASSCF calculation is sufficient to prove our concept of $\heff$ and correlating it with the errors of bond dissociation energies (BDEs). All of our calculations were performed using PySCF \cite{Sun151664,Sun181759,Sun20024109} with the def2-QZVPPD basis set.\cite{Weigend0312753} 
Our data set consists of main-group compounds in the BDE99 data set \cite{Mardirossian172315,Karton11165} and transition-metal (TM) diatomic molecules 
including oxides \cite{Moltved1918432}, 
hydrides \cite{Moltved183479},
nitrides and dimers \cite{Furche060021}.
In total, 145 molecules with 172 types of bond-breaking processes are considered.
For some of the molecules such as HOCl, data are available for two types of bond-breaking processes, namely H-OCl and HO-Cl.

In \Fig{dBDEhbarLDA}, we supplement the BDE errors for the local density approximation (LDA) functional, \cite{Vosko801200}
and show its correlation with $\heff$. 
The LDA results are similar to PBE shown in the main text; typical weakly and strongly correlated compounds have been enclosed by the black and blue box, respectively.
Here the general trend is that data points are dispersed in a band region enclosed by the green reference lines, suggesting a stronger correlation between the LDA error and $\heff$.

In Table S1, we further supplement the BDE values of PBE, LDA and the reference along with $\heff$ for the molecules tested. 
For a bond dissociation in A--B, the DFA result is calculated by
\begin{align}
	E_{\text{BDE}}^{\text{DFA}} \equiv E^{\text{DFA}}(\text{A}) + E^{\text{DFA}}(\text{B}) - E^{\text{DFA}}(\text{A--B}).
\end{align}
The reference BDE values are either extracted from high-level calculations \cite{Mardirossian172315} or experimental data \cite{Moltved183479,Moltved1918432,Furche060021}, 
where in the later case all the energy corrections beyond the electronic Schr\"odinger equation, including the zero-point energies (ZPEs), the relativistic corrections, etc., have been deducted in order to compare the electronic energy.

\setlength{\tabcolsep}{4mm}{
\LTcapwidth=\textwidth
\begin{longtable}{|c|c|c|c|c|c|c|}
	\caption{
		Supplementary data of bond dissociation energies (BDEs), signed errors of DFAs (in kcal/mol) and $\heff$ of 145 molecules with 172 types of bond-breaking. 
		For some molecules, our LDA calculations fail to converge. In such cases, we use the literature data when available (in italic) and leave blank otherwise.
	} 
	\\
	\hline Compound & Reference & PBE & PBE error & LDA & LDA error & $\heff$ \\
	\hline
	\endfirsthead
	\multicolumn{7}{l}{{\tablename\ \thetable. ({\it Continued. })}} \\
	\hline 
	Compound & Reference & PBE & PBE error & LDA & LDA error & $\heff$ \\
	\hline
	\endhead
	\hline 
	\endfoot
	\hline 
	\endlastfoot
	\hline
	\multicolumn{7}{|c|}{Nonmultireference subset of BDE99} \\
	\hline
	H-OCl   &  100.8  \cite{Mardirossian172315}  &  94.0   & -6.8 &  {\it  106.6  } \cite{Mardirossian172315}  & +5.8 &  0.73  \\
	HO-Cl   &  59.0  \cite{Mardirossian172315}  &  65.5   & +6.5 &  {\it  86.9  } \cite{Mardirossian172315}  & +27.9 &  0.73  \\
	HO-F  &  51.5  \cite{Mardirossian172315}  &  62.4   & +10.9 &   87.1    & +35.7 &  0.59  \\
	O-NO  &  75.1  \cite{Mardirossian172315}  &  97.7   & +22.6 &  {\it  124.8  } \cite{Mardirossian172315}  & +49.7 &  0.49  \\
	N-NO  &  118.1  \cite{Mardirossian172315}  &  138.6   & +20.5 &  {\it  162.9  } \cite{Mardirossian172315}  & +44.8 &  0.65  \\
	NN-O  &  42.4  \cite{Mardirossian172315}  &  67.0   & +24.7 &   94.2    & +51.8 &  0.65  \\
	H-ONO  &  84.8  \cite{Mardirossian172315}  &  75.9   & -8.9 &   85.4    & +0.6 &  0.58  \\
	HO-NO  &  52.7  \cite{Mardirossian172315}  &  63.7   & +11.0 &  {\it  86.1  } \cite{Mardirossian172315}  & +33.4 &  0.58  \\
	H-NC  &  116.9  \cite{Mardirossian172315}  &  113.9   & -3.0 &   127.3    & +10.5 &  0.72  \\
	HN-C  &  215.1  \cite{Mardirossian172315}  &  222.9   & +7.8 &   251.2    & +36.1 &  0.72  \\
	HO-NC  &  61.6  \cite{Mardirossian172315}  &  69.2   & +7.6 &   92.7    & +31.1 &  0.69  \\
	HN-CO  &  91.9  \cite{Mardirossian172315}  &  107.0   & +15.1 &   131.8    & +39.9 &  0.68  \\
	HC-NO  &  128.0  \cite{Mardirossian172315}  &  144.2   & +16.2 &  {\it  172.7  } \cite{Mardirossian172315}  & +44.7 &  0.60  \\
	H-NO  &  53.1  \cite{Mardirossian172315}  &  51.2   & -1.9 &  {\it  62.1  } \cite{Mardirossian172315}  & +9.0 &  0.55  \\
	HN-O  &  122.8  \cite{Mardirossian172315}  &  134.8   & +12.0 &   165.3    & +42.5 &  0.55  \\
	HO-CH$_3$  &  98.4  \cite{Mardirossian172315}  &  100.1   & +1.7 &   123.0    & +24.6 &  0.96  \\
	HC-OH  &  131.1  \cite{Mardirossian172315}  &  137.7   & +6.7 &   163.4    & +32.3 &  0.87  \\
	HCO-H  &  43.1  \cite{Mardirossian172315}  &  37.1   & -5.9 &   47.1    & +4.0 &  0.87  \\
	HCO-OH  &  115.3  \cite{Mardirossian172315}  &  117.1   & +1.8 &   140.5    & +25.2 &  0.74  \\
	H$_2$CC-O  &  173.5  \cite{Mardirossian172315}  &  188.5   & +15.0 &   216.3    & +42.8 &  0.62  \\
	H$_2$C-CO  &  83.0  \cite{Mardirossian172315}  &  94.6   & +11.6 &   115.7    & +32.7 &  0.62  \\
	OHC-CHO  &  76.3  \cite{Mardirossian172315}  &  72.6   & -3.6 &   86.8    & +10.5 &  0.61  \\
	H$_3$C-CHO  &  90.6  \cite{Mardirossian172315}  &  89.2   & -1.3 &   106.5    & +16.0 &  0.69  \\
	F$_2$C-O  &  161.9  \cite{Mardirossian172315}  &  171.1   & +9.2 &   199.1    & +37.2 &  0.74  \\
	OHC-F  &  124.3  \cite{Mardirossian172315}  &  128.4   & +4.1 &   152.5    & +28.2 &  0.71  \\
	OHC-H  &  95.2  \cite{Mardirossian172315}  &  90.7   & -4.6 &   100.8    & +5.5 &  0.66  \\
	H$_2$C-O  &  183.9  \cite{Mardirossian172315}  &  191.3   & +7.4 &   220.7    & +36.8 &  0.66  \\
	H-CO  &  19.7  \cite{Mardirossian172315}  &  26.3   & +6.6 &   34.0    & +14.3 &  0.61  \\
	HC-O  &  195.2  \cite{Mardirossian172315}  &  210.5   & +15.3 &   240.4    & +45.2 &  0.61  \\
	SC-S  &  108.6  \cite{Mardirossian172315}  &  121.6   & +13.1 &   141.6    & +33.0 &  0.59  \\
	OC-O  &  130.4  \cite{Mardirossian172315}  &  147.1   & +16.7 &   174.0    & +43.6 &  0.72  \\
	H$_2$C-NH$_2$  &  108.9  \cite{Mardirossian172315}  &  113.1   & +4.1 &   135.6    & +26.7 &  0.98  \\
	H$_2$CNH-H  &  42.8  \cite{Mardirossian172315}  &  44.3   & +1.4 &   52.4    & +9.6 &  0.98  \\
	H$_3$C-NH  &  83.7  \cite{Mardirossian172315}  &  89.0   & +5.3 &   108.6    & +24.9 &  0.93  \\
	H-CH$_2$NH  &  35.2  \cite{Mardirossian172315}  &  35.4   & +0.2 &   39.2    & +4.0 &  0.93  \\
	H-CH$_2$NH$_2$  &  100.0  \cite{Mardirossian172315}  &  94.6   & -5.4 &   104.5    & +4.5 &  1.00  \\
	CH$_3$NH-H  &  107.7  \cite{Mardirossian172315}  &  103.4   & -4.2 &   117.8    & +10.1 &  1.00  \\
	H-CHNH  &  103.2  \cite{Mardirossian172315}  &  97.8   & -5.3 &   107.7    & +4.5 &  0.62  \\
	HC-NH  &  168.9  \cite{Mardirossian172315}  &  181.0   & +12.1 &   208.5    & +39.5 &  0.57  \\
	HCN-H  &  22.8  \cite{Mardirossian172315}  &  27.9   & +5.1 &   35.6    & +12.7 &  0.57  \\
	H-CN  &  132.1  \cite{Mardirossian172315}  &  128.8   & -3.3 &   141.3    & +9.2 &  0.65  \\
	HC-N  &  229.2  \cite{Mardirossian172315}  &  241.6   & +12.4 &   268.2    & +39.0 &  0.65  \\
	NC-CN  &  139.3  \cite{Mardirossian172315}  &  143.5   & +4.1 &   161.1    & +21.8 &  0.53  \\
	H-SH  &  96.2  \cite{Mardirossian172315}  &  94.1   & -2.1 &   106.8    & +10.6 &  0.84  \\
	HOO-H  &  93.6  \cite{Mardirossian172315}  &  87.0   & -6.5 &   99.4    & +5.9 &  0.62  \\
	HO-OH  &  54.7  \cite{Mardirossian172315}  &  62.3   & +7.6 &   86.7    & +32.0 &  0.62  \\
	S-SH  &  77.4  \cite{Mardirossian172315}  &  85.5   & +8.1 &   103.0    & +25.6 &  0.79  \\
	SS-H  &  60.9  \cite{Mardirossian172315}  &  58.2   & -2.6 &   67.4    & +6.5 &  0.79  \\
	O-OH  &  68.3  \cite{Mardirossian172315}  &  85.1   & +16.8 &   111.3    & +43.0 &  0.66  \\
	OO-H  &  54.7  \cite{Mardirossian172315}  &  51.4   & -3.3 &   60.7    & +6.0 &  0.66  \\
	H-OH  &  125.8  \cite{Mardirossian172315}  &  124.4   & -1.4 &   142.3    & +16.6 &  0.95  \\
	HN-NN  &  20.2  \cite{Mardirossian172315}  &  41.0   & +20.8 &   65.7    & +45.5 &  0.64  \\
	H$_2$N-NH$_2$  &  73.1  \cite{Mardirossian172315}  &  75.7   & +2.6 &   99.7    & +26.6 &  0.87  \\
	HN-NH  &  130.3  \cite{Mardirossian172315}  &  137.7   & +7.4 &   166.3    & +35.9 &  0.56  \\
	N-NH  &  141.8  \cite{Mardirossian172315}  &  158.8   & +17.0 &   183.5    & +41.7 &  0.54  \\
	H$_2$N-Cl  &  65.5  \cite{Mardirossian172315}  &  69.7   & +4.3 &  {\it  89.8  } \cite{Mardirossian172315}  & +24.3 &  0.77  \\
	H$_2$N-H  &  115.4  \cite{Mardirossian172315}  &  113.3   & -2.1 &   129.4    & +13.9 &  0.97  \\
	HN-H  &  99.5  \cite{Mardirossian172315}  &  100.2   & +0.7 &   112.5    & +13.0 &  0.93  \\
	H$_2$CC-CH$_2$  &  153.4  \cite{Mardirossian172315}  &  159.8   & +6.4 &   181.0    & +27.6 &  0.58  \\
	HCC-CH$_3$  &  131.6  \cite{Mardirossian172315}  &  133.9   & +2.3 &   153.1    & +21.6 &  0.65  \\
	H$_2$CCH-CH$_3$  &  107.6  \cite{Mardirossian172315}  &  105.7   & -1.9 &   124.2    & +16.5 &  0.58  \\
	H$_2$CCH-F  &  127.8  \cite{Mardirossian172315}  &  131.9   & +4.1 &   156.6    & +28.8 &  1.05  \\
	H$_3$C-CH$_3$  &  97.3  \cite{Mardirossian172315}  &  96.8   & -0.6 &   115.6    & +18.2 &  1.02  \\
	H$_2$C-CH$_2$  &  182.6  \cite{Mardirossian172315}  &  182.5   & -0.1 &   206.7    & +24.1 &  0.60  \\
	H$_2$CC-H  &  86.2  \cite{Mardirossian172315}  &  88.4   & +2.2 &   96.1    & +9.9 &  1.05  \\
	H-HCCH  &  40.6  \cite{Mardirossian172315}  &  43.0   & +2.5 &   47.2    & +6.6 &  1.05  \\
	H$_2$C-C  &  169.2  \cite{Mardirossian172315}  &  175.0   & +5.8 &   198.1    & +28.9 &  0.65  \\
	HCC-F  &  132.3  \cite{Mardirossian172315}  &  141.2   & +8.9 &   167.6    & +35.3 &  0.66  \\
	HC-CF  &  181.5  \cite{Mardirossian172315}  &  188.6   & +7.0 &   216.7    & +35.1 &  0.66  \\
	HCC-H  &  139.4  \cite{Mardirossian172315}  &  137.8   & -1.5 &   150.6    & +11.2 &  0.65  \\
	HC-CH  &  237.1  \cite{Mardirossian172315}  &  245.5   & +8.4 &   275.1    & +38.0 &  0.65  \\
	H$_3$C-F  &  115.1  \cite{Mardirossian172315}  &  119.1   & +4.0 &   143.3    & +28.2 &  0.94  \\
	H$_3$C-H  &  112.6  \cite{Mardirossian172315}  &  110.0   & -2.5 &   123.2    & +10.6 &  1.02  \\
	H$_2$C-H  &  117.1  \cite{Mardirossian172315}  &  115.4   & -1.7 &   126.4    & +9.2 &  0.96  \\
	HC-H  &  106.5  \cite{Mardirossian172315}  &  109.8   & +3.3 &   120.4    & +13.9 &  0.92  \\
	FC-F  &  126.1  \cite{Mardirossian172315}  &  131.3   & +5.2 &   156.7    & +30.6 &  0.93  \\
	Cl-CN  &  104.1  \cite{Mardirossian172315}  &  109.2   & +5.1 &  {\it  127.7  } \cite{Mardirossian172315}  & +23.6 &  0.65  \\
	H$_3$B-BH$_3$  &  44.5  \cite{Mardirossian172315}  &  51.4   & +7.0 &   66.7    & +22.2 &  1.07  \\
	SS-O  &  104.5  \cite{Mardirossian172315}  &  113.3   & +8.7 &   140.5    & +36.0 &  0.70  \\
	S-SO  &  82.3  \cite{Mardirossian172315}  &  87.4   & +5.1 &   108.3    & +26.0 &  0.70  \\
	O$_2$S-O  &  86.3  \cite{Mardirossian172315}  &  94.4   & +8.0 &   119.9    & +33.6 &  0.76  \\
	OS-O  &  134.2  \cite{Mardirossian172315}  &  139.2   & +5.1 &   168.3    & +34.1 &  0.60  \\
	PP-PP  &  55.4  \cite{Mardirossian172315}  &  65.7   & +10.4 &   78.8    & +23.5 &  0.74  \\
	\hline
	\multicolumn{7}{|c|}{Multireference subset of BDE99} \\
	\hline
	F-OOF  &  17.7  \cite{Mardirossian172315}  &  31.1   & +13.5 &   51.8    & +34.2 &  0.42  \\
	FO-OF  &  46.2  \cite{Mardirossian172315}  &  62.3   & +16.0 &   82.0    & +35.7 &  0.42  \\
	O-ClO  &  62.7  \cite{Mardirossian172315}  &  80.0   & +17.3 &  {\it  108.2  } \cite{Mardirossian172315}  & +45.6 &  0.62  \\
	ClO-O  &  60.9  \cite{Mardirossian172315}  &  79.8   & +18.9 &  {\it  100.5  } \cite{Mardirossian172315}  & +39.5 &  0.26  \\
	Cl-OO  &  5.6  \cite{Mardirossian172315}  &  17.6   & +12.1 &  {\it  30.2  } \cite{Mardirossian172315}  & +24.6 &  0.26  \\
	FO-O  &  81.7  \cite{Mardirossian172315}  &  103.0   & +21.3 &   127.0    & +45.3 &  0.36  \\
	F-OO  &  13.9  \cite{Mardirossian172315}  &  31.2   & +17.3 &   49.1    & +35.2 &  0.36  \\
	ClO-Cl  &  36.0  \cite{Mardirossian172315}  &  38.0   & +2.0 &  {\it  54.6  } \cite{Mardirossian172315}  & +18.6 &  0.69  \\
	FO-F  &  40.7  \cite{Mardirossian172315}  &  51.5   & +10.8 &   74.0    & +33.3 &  0.53  \\
	H-OF  &  105.6  \cite{Mardirossian172315}  &  100.4   & -5.2 &   114.4    & +8.8 &  0.59  \\
	HOO-O  &  57.8  \cite{Mardirossian172315}  &  76.8   & +19.0 &   99.2    & +41.5 &  0.44  \\
	HO-OO  &  5.3  \cite{Mardirossian172315}  &  18.3   & +13.0 &   35.8    & +30.6 &  0.44  \\
	H-OOO  &  85.9  \cite{Mardirossian172315}  &  86.9   & +1.1 &   92.9    & +7.0 &  0.44  \\
	OO-O  &  26.6  \cite{Mardirossian172315}  &  41.3   & +14.7 &   67.0    & +40.4 &  0.25  \\
	SSS-S  &  66.0  \cite{Mardirossian172315}  &  74.0   & +8.0 &   92.0    & +26.0 &  0.20  \\
	SS-SS  &  25.9  \cite{Mardirossian172315}  &  28.8   & +3.0 &   46.4    & +20.5 &  0.20  \\
	\hline
	\multicolumn{7}{|c|}{Transition-metal compounds} \\
	\hline
	ScO  &  161.9  \cite{Moltved1918432}  &  184.4   & +22.6 &   209.4    & +47.6 &  0.75  \\
	TiO  &  160.4  \cite{Moltved1918432}  &  188.1   & +27.7 &   215.3    & +54.9 &  0.60  \\
	VO  &  153.5  \cite{Moltved1918432}  &  183.1   & +29.6 &      &  &  0.51  \\
	CrO  &  106.7  \cite{Moltved1918432}  &  120.8   & +14.0 &      &  &  0.45  \\
	MnO  &  88.4  \cite{Moltved1918432}  &  126.5   & +38.1 &   148.9    & +60.5 &  0.37  \\
	FeO  &  99.7  \cite{Moltved1918432}  &  128.5   & +28.8 &      &  &  0.34  \\
	CoO  &  91.6  \cite{Moltved1918432}  &  115.1   & +23.4 &      &  &  0.35  \\
	NiO  &  86.1  \cite{Moltved1918432}  &  112.8   & +26.7 &      &  &  0.35  \\
	CuO  &  66.6  \cite{Moltved1918432}  &  77.9   & +11.3 &   95.5    & +28.9 &  0.34  \\
	ZnO  &  39.8  \cite{Moltved1918432}  &  46.9   & +7.2 &   66.3    & +26.5 &  0.52  \\
	YO  &  171.1  \cite{Moltved1918432}  &  181.7   & +10.6 &   206.5    & +35.4 &  0.78  \\
	ZrO  &  184.1  \cite{Moltved1918432}  &  198.9   & +14.8 &   224.4    & +40.3 &  0.75  \\
	NbO  &  174.8  \cite{Moltved1918432}  &  192.1   & +17.2 &      &  &  0.70  \\
	MoO  &  126.7  \cite{Moltved1918432}  &  141.0   & +14.2 &      &  &  0.51  \\
	TcO  &  131.2  \cite{Moltved1918432}  &  146.0   & +14.8 &   169.6    & +38.4 &  0.51  \\
	RuO  &  127.0  \cite{Moltved1918432}  &  134.8   & +7.7 &   162.7    & +35.7 &  0.50  \\
	RhO  &  98.3  \cite{Moltved1918432}  &  121.1   & +22.8 &   151.9    & +53.6 &  0.46  \\
	PdO  &  55.7  \cite{Moltved1918432}  &  81.3   & +25.6 &   102.6    & +46.8 &  0.43  \\
	AgO  &  52.8  \cite{Moltved1918432}  &  55.9   & +3.1 &   71.3    & +18.5 &  0.84  \\
	CdO  &  23.1  \cite{Moltved1918432}  &  29.0   & +5.9 &   38.9    & +15.8 &  0.79  \\
	HfO  &  196.4  \cite{Moltved1918432}  &  205.7   & +9.3 &   235.2    & +38.7 &  0.66  \\
	TaO  &  202.4  \cite{Moltved1918432}  &  200.0   & -2.4 &   228.7    & +26.4 &  0.57  \\
	WO  &  169.0  \cite{Moltved1918432}  &  185.4   & +16.4 &   217.4    & +48.4 &  0.52  \\
	OsO  &  133.1  \cite{Moltved1918432}  &  138.8   & +5.7 &      &  &  0.48  \\
	IrO  &  106.3  \cite{Moltved1918432}  &  130.3   & +24.0 &      &  &  0.51  \\
	PtO  &  94.6  \cite{Moltved1918432}  &  116.6   & +22.0 &      &  &  0.51  \\
	AuO  &  51.5  \cite{Moltved1918432}  &  63.9   & +12.4 &   84.0    & +32.5 &  0.72  \\
	ScO$^+$  &  165.4  \cite{Moltved1918432}  &  180.2   & +14.7 &   209.1    & +43.7 &  0.67  \\
	TiO$^+$  &  158.3  \cite{Moltved1918432}  &  172.6   & +14.3 &   200.8    & +42.4 &  0.64  \\
	VO$^+$  &  136.9  \cite{Moltved1918432}  &  156.3   & +19.4 &      &  &  0.50  \\
	CrO$^+$  &  82.4  \cite{Moltved1918432}  &  99.1   & +16.7 &   124.6    & +42.2 &  0.39  \\
	MnO$^+$  &  72.6  \cite{Moltved1918432}  &  83.9   & +11.3 &      &  &  0.27  \\
	FeO$^+$  &  83.0  \cite{Moltved1918432}  &  111.3   & +28.3 &      &  &  0.39  \\
	CoO$^+$  &  70.6  \cite{Moltved1918432}  &  95.9   & +25.3 &   121.4    & +50.8 &  0.33  \\
	NiO$^+$  &  42.3  \cite{Moltved1918432}  &  84.5   & +42.2 &      &  &  0.38  \\
	CuO$^+$  &  26.4  \cite{Moltved1918432}  &  48.9   & +22.5 &   63.9    & +37.5 &  1.07  \\
	ZnO$^+$  &  39.2  \cite{Moltved1918432}  &  48.5   & +9.3 &   62.1    & +22.9 &  0.77  \\
	YO$^+$  &  172.1  \cite{Moltved1918432}  &  189.3   & +17.2 &   208.9    & +36.8 &  0.75  \\
	ZrO$^+$  &  180.7  \cite{Moltved1918432}  &  189.1   & +8.3 &   215.6    & +34.9 &  0.73  \\
	NbO$^+$  &  165.1  \cite{Moltved1918432}  &  185.6   & +20.5 &      &  &  0.67  \\
	MoO$^+$  &  117.1  \cite{Moltved1918432}  &  143.1   & +26.1 &   173.2    & +56.1 &  0.51  \\
	RuO$^+$  &  89.2  \cite{Moltved1918432}  &  111.0   & +21.8 &   127.7    & +38.5 &  0.43  \\
	RhO$^+$  &  72.9  \cite{Moltved1918432}  &  96.7   & +23.9 &      &  &  0.41  \\
	PdO$^+$  &  35.2  \cite{Moltved1918432}  &  74.4   & +39.2 &      &  &  0.44  \\
	AgO$^+$  &  29.0  \cite{Moltved1918432}  &  24.1   & -4.9 &   37.8    & +8.8 &  1.05  \\
	HfO$^+$  &  180.2  \cite{Moltved1918432}  &  185.1   & +5.0 &   213.9    & +33.7 &  0.72  \\
	TaO$^+$  &  188.7  \cite{Moltved1918432}  &  188.4   & -0.3 &   218.3    & +29.6 &  0.67  \\
	WO$^+$  &  169.8  \cite{Moltved1918432}  &  183.8   & +14.0 &      &  &  0.62  \\
	ReO$^+$  &  100.3  \cite{Moltved1918432}  &  120.3   & +20.0 &      &  &  0.52  \\
	OsO$^+$  &  95.6  \cite{Moltved1918432}  &  125.7   & +30.1 &      &  &  0.33  \\
	IrO$^+$  &  104.8  \cite{Moltved1918432}  &  122.9   & +18.1 &   151.1    & +46.3 &  0.50  \\
	PtO$^+$  &  78.8  \cite{Moltved1918432}  &  104.0   & +25.3 &   132.7    & +53.9 &  0.48  \\
	AuO$^+$  &  18.6  \cite{Moltved1918432}  &  54.0   & +35.4 &   76.0    & +57.5 &  0.64  \\
	ScH  &  51.4  \cite{Moltved183479}  &  54.4   & +3.1 &   64.3    & +12.9 &  0.51  \\
	TiH  &  53.6  \cite{Moltved183479}  &  65.6   & +12.1 &   73.9    & +20.4 &  0.88  \\
	VH  &  54.6  \cite{Moltved183479}  &  74.1   & +19.5 &  {\it  82.7  } \cite{Moltved183479}  & +28.1 &  0.77  \\
	CrH  &  46.2  \cite{Moltved183479}  &  52.5   & +6.3 &   60.4    & +14.2 &  0.64  \\
	MnH  &  34.2  \cite{Moltved183479}  &  45.3   & +11.1 &   49.8    & +15.5 &  0.90  \\
	FeH  &  43.2  \cite{Moltved183479}  &  56.2   & +12.9 &  {\it  66.6  } \cite{Moltved183479}  & +23.4 &  0.55  \\
	CoH  &  53.4  \cite{Moltved183479}  &  60.7   & +7.4 &   72.8    & +19.5 &  0.38  \\
	NiH  &  62.1  \cite{Moltved183479}  &  66.7   & +4.6 &  {\it  78.4  } \cite{Moltved183479}  & +16.3 &  0.68  \\
	CuH  &  60.6  \cite{Moltved183479}  &  64.6   & +4.0 &   75.6    & +15.0 &  0.72  \\
	ZnH  &  23.9  \cite{Moltved183479}  &  23.0   & -0.9 &   29.3    & +5.5 &  0.82  \\
	Ti$_2$  &  37.3  \cite{Furche060021}  &  77.2   & +39.9 &   108.2    & +70.9 &  0.41  \\
	V$_2^+$  &  73.2  \cite{Furche060021}  &  105.9   & +32.7 &  {\it  129.0  } \cite{Furche060021}  & +55.8 &  0.35  \\
	V$_2$  &  65.8  \cite{Furche060021}  &  105.4   & +39.6 &  {\it  113.1  } \cite{Furche060021}  & +47.3 &  0.35  \\
	Cr$_2$  &  35.3  \cite{Furche060021}  &  24.6   & -10.7 &   62.7    & +27.4 &  0.20  \\
	Fe$_2$  &  27.8  \cite{Furche060021}  &  57.3   & +29.6 &  {\it  88.8  } \cite{Furche060021}  & +61.1 &  0.19  \\
	Cu$_2$  &  47.5  \cite{Furche060021}  &  49.0   & +1.5 &   60.1    & +12.7 &  0.75  \\
	ScN  &  115.3  \cite{Furche060021}  &  123.8   & +8.5 &   143.7    & +28.4 &  0.54  \\
	TiN  &  127.0  \cite{Furche060021}  &  144.6   & +17.7 &   170.8    & +43.9 &  0.55  \\
	VN  &  120.0  \cite{Furche060021}  &  147.1   & +27.2 &  {\it  157.5  } \cite{Furche060021}  & +37.6 &  0.49  \\
	CrN  &  100.0  \cite{Furche060021}  &  106.3   & +6.3 &   127.0    & +27.0 &  0.42  \\
\end{longtable}
}


\section{Some details of DFA calculations in the semiclassical limit}

By the energy scaling argument in the main text, the energy functional $(E_\hbar)$ for $\hbar \neq 1$ and for $\hbar = 1$ ($E_1$) are related by the following equation,
\begin{align}
	E_\hbar[\rho_\hbar] = \frac{1}{\hbar^2} E_1 [\rho_1]. \label{EhE1}
\end{align}
Here $\rho_\hbar$ is an arbitrary $v$-representable density for the problem $\hbar\neq 1$, and $\rho_1(\br) = \hbar^6 \rho_\hbar(\hbar^2 \br)$ is the scaled density. \Eq{EhE1} can be viewed as the defining equation for $E_\hbar$. 

By the Kohn-Sham decomposition, 
\begin{align}
	E_1[\rho_1] = T_{s,1}[\rho_1] + V_{\rm ext,1}[\rho_1]+J_{1}[\rho_1] + E_{xc,1}[\rho_1]. \label{KShbar}
\end{align}
Let us analyze terms on the RHS one by one. 
\begin{align}
	& V_{\rm ext,1}[\rho_1] =  \int \hbar^6 \rho_\hbar(\hbar^2 \br) V_{\rm ext,1}(\br) d\br = \hbar^2\int \rho_\hbar(\br') V_{\rm ext,\hbar}(\br') d\br'. \label{Vhbar}
\end{align}
Here we have used $V_{\rm ext, 1}(\br) = \hbar^2 V_{\rm ext, \hbar}(\hbar^2 \br)$ and change of variable $\br' = \hbar^2 \br$. By the same argument, 
\begin{align}
	& J_{1}[\rho_1] = \frac{1}{2}\int \frac{\rho_1(\br)\rho_1(\br')}{|\br-\br'|} d\br d\br' = \frac{1}{2}\int \frac{\hbar^6 \rho_\hbar(\hbar^2 \br)\hbar^6 \rho_\hbar(\hbar^2 \br')}{|\br-\br'|} d\br d\br' = \frac{1}{2}\hbar^2\int \frac{\rho_\hbar( \br)\rho_\hbar(\br')}{|\br-\br'|} d\br d\br'. \label{Jhbar}
\end{align}
The Kohn-Sham kinetic energy calculation is slightly more complicated because it involves the Kohn-Sham orbitals. However, because the problem of $\hbar\neq 1$ and $\hbar = 1$ are equivalent up to a scaling of coordinate, the corresponding Kohn-Sham orbitals shall also have similar relations, i.e., $\phi_{i,1}(\br) = \hbar^3\phi_{i,\hbar}(\hbar^2 \br)$.  
It follows that
\begin{align}
	T_{s,1}[\rho_1] &=  -\frac{1}{2}\sum_{i}^{\rm occ.}\int \phi_{i,1}^*(\br)\nabla^2\phi_{i,1}(\br)d\br = -\frac{\hbar^6}{2}\sum_{i}^{\rm occ.}\int \phi_{i,\hbar}^*(\hbar^2 \br)\nabla^2\phi_{i,\hbar}(\hbar^2\br)d\br \nl
	&= -\frac{\hbar^2}{2}\sum_{i}^{\rm occ.}\int \phi_{i,\hbar}^*( \br)\nabla^2\phi_{i,\hbar}(\br)d\br. \label{Tshbar}
\end{align}
Substituting \Eqs{Vhbar}--\eqref{Tshbar} into \Eq{KShbar} and comparing with \Eq{EhE1}, we have
\begin{align}
	E_\hbar[\rho_\hbar] &= -\frac{1}{2}\sum_{i}^{\rm occ.}\int \phi_{i,\hbar}^*( \br)\nabla^2\phi_{i,\hbar}(\br)d\br + \int \rho_\hbar(\br') V_{\rm ext,1}(\br') d\br'+\frac{1}{2}\int \frac{\rho_\hbar( \br)\rho_\hbar(\br')}{|\br-\br'|} d\br d\br' + \frac{1}{\hbar^2}E_{xc,1}[\rho_1] \nl 
	&= 	T_{s,1}[\rho_\hbar] + V_{\rm ext,1}[\rho_\hbar] + J_{1}[\rho_\hbar]+ \frac{1}{\hbar^2}E_{xc,1}[\rho_1] .
\end{align}
This suggests that in the evaluation of $E_\hbar$ for a given density, we only have to plug the density directly into the functional form of $E_1$ except the xc functional, which should be replaced by
\begin{align}
	E_{xc,\hbar}[\rho_\hbar] =  \frac{1}{\hbar^2} E_{xc,1}[\rho_{1}]. \label{Exchbar}
\end{align}
\Eq{Exchbar} shall be viewed as the defining equation for the xc functional for $\hbar\neq 1$. This is true for the exact xc functional as well as DFAs.
In addition, for most of the mainstream DFA exchange as well as the exact exchange functional, one can show that $\frac{1}{\hbar^2} E_{x,1}[\rho_{1}]=E_{x,1}[\rho_\hbar]$ using our scaling arguments above or the uniform scaling equality \cite{Levy852010}. This further reduces the re-scaling definition to only the correlation functional,
\begin{align}
	E_{xc,\hbar}[\rho_\hbar] = E_{x,1}[\rho_\hbar] + \frac{1}{\hbar^2} E_{c,1}[\rho_{1}],
\end{align}

As a side remark, \Eq{Exchbar} can also be derived from the constrained search definition of the exact functional \cite{Levy796062}, which is in similar spirit to the derivation of the scaling relation with respect to the interaction strength  $\lambda$\cite{Parr94}.


\section{Some details of the asymptotic leading term of LDA}


\begin{figure}[htpb]
	\begin{center}
		\includegraphics[width=0.8\linewidth]{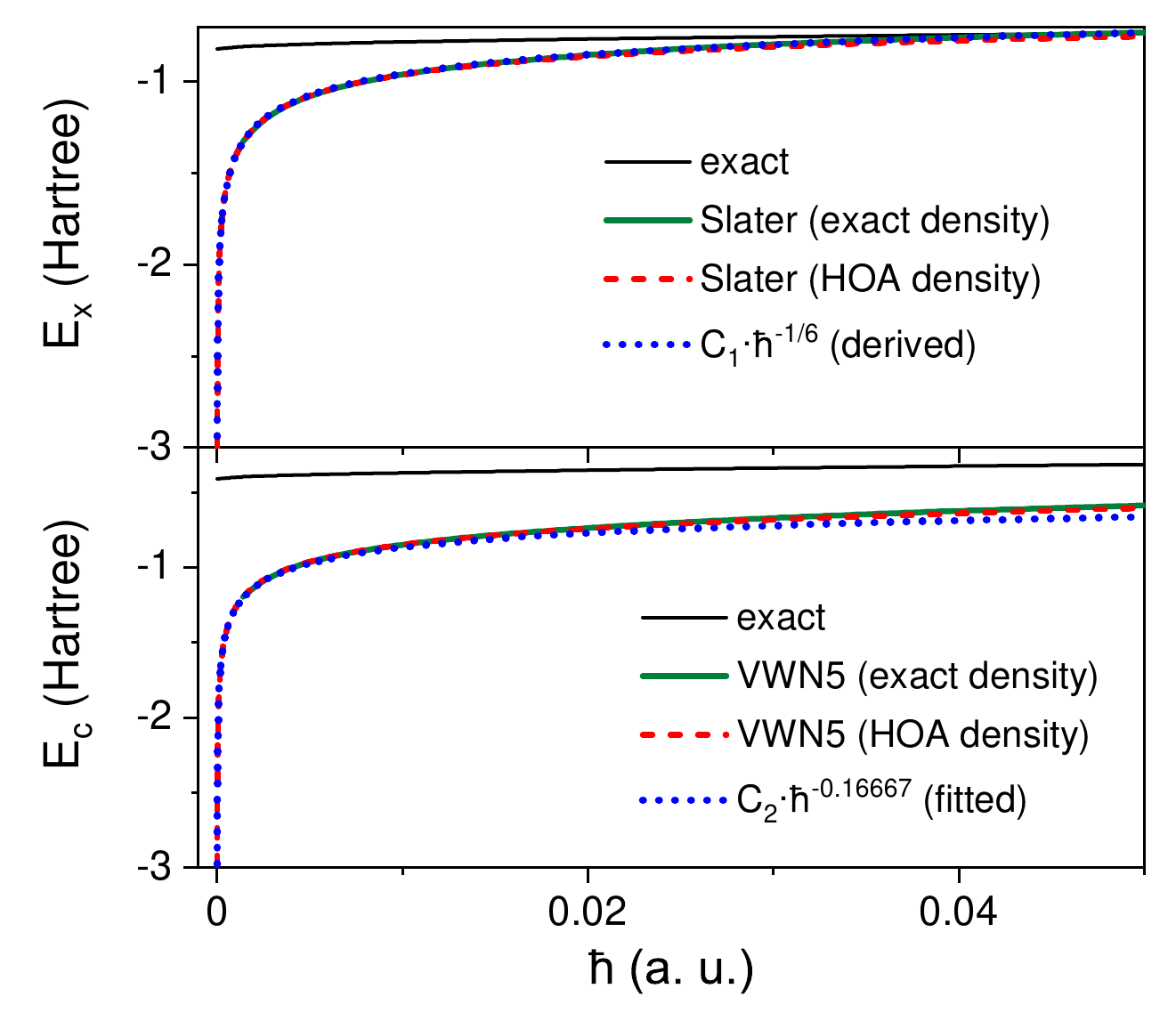}
	\end{center}
	\caption{
		LDA exchange (Slater) and correlation (VWN5) energy for harmonium as $\hbar \rightarrow 0$, in comparison with the exact results. The blue dotted curves show the asymptotic behavior of LDA.
	} \label{HookesHeASYM}
\end{figure}

Here we give a brief derivation of the asymptotic $\mO(\hbar^{-1/6})$ behavior of the LDA exchange energy in the limit $\hbar \rightarrow 0$ for model atoms. 
Taking harmonium with $\lambda=1$ and arbitrary $\omega$ for an example, the ground state density under HOA can be written as 
\begin{align}
	\rho_{\text{HOA}} = \frac{B}{\sqrt h} \exp\Big[ -\frac{a}{\hbar}(r - r_0)^2 \Big]. \label{rhoA}
\end{align}
Here $B= 2 \times \frac{1}{4\pi r_0^2} \sqrt{\frac{a}{\pi}}$, $a = (3-\sqrt{3})\omega$ and $r_0 = (2\omega)^{-2/3}$.
As the main text shows, $\rho_{\rm HOA}$ is an extremely good approximation to the exact density in approaching the semiclassical limit.
Substituting $\rho_{\rm HOA}$ into the LDA (Slater) exchange functional, we can obtain that
\begin{align}
	E_x^{\rm LDA}[\rho_{\text{HOA}}] = & - C_x \int \rho_{\text{HOA}}^{\frac{4}{3}} d\br = - \frac{4\pi C_x B^\frac{4}{3}}{\hbar^{2/3}}  \int_0^\infty  \exp\Big[ -\frac{4a}{3\hbar}(r - r_0)^2 \Big]  r^2 dr .
\end{align}
As $\hbar \rightarrow 0$, the integrand contains an extremely narrow Gaussian centered in the positive $r$ region, which can be well approximated by
\begin{align}
	\int_0^\infty  \exp\Big[ -\frac{4a}{3\hbar}(r - r_0)^2 \Big]  r^2 dr \approx \int_{-\infty}^\infty  \exp\Big[ -\frac{4a}{3\hbar}(r - r_0)^2 \Big]  r_0^2 dr = r_0^2 \sqrt{\frac{3\pi\hbar}{4a}}.
\end{align}
Therefore, 
\begin{align}
	E_x^{\rm LDA}[\rho_{\text{HOA}}] \approx  -\mO\Big(\hbar^{-\frac{1}{6}}\Big). \label{ExA}
\end{align}
The VWN5 correlation has more complicated functional form, which is difficult to analytically determine the leading-order divergence. Nevertheless, by fitting the numerical results with small $\hbar$'s, we find that it has similar $\mO(\hbar^{-1/6})$ behavior.

The above conclusions have been verified in 
\Fig{HookesHeASYM}. Substituting the exact or HOA density into the LDA exchange (and also correlation) functional essentially leads to the same result for small $\hbar$, and they all agree with the $\mO(\hbar^{-1/6})$ asymptotic behavior, strongly deviating from the exact curves.
Here for 2-electron spin-unpolarized problem, the exact exchange has simple relation with the Hartree energy as $E_{x}^{\rm exact} = -\frac{1}{2}J$. The exact correlation energy can be then calculated by $E_{c}^{\rm exact} = E_{xc}^{\rm exact} - E_{x}^{\rm exact}$.

For 2-electron diatomic molecular models, similarly, the ground state density can be approximated by the following HOA expression written in cylindrical coordinates,
\begin{align}
	\rho_{\text{mol}} \approx \frac{A}{\hbar}\exp\Big[ -\frac{b}{\hbar}z^2 - \frac{c}{\hbar}(s-s_0)^2 \Big]. \label{rhoM}
\end{align}
Here $s$ and $z$ are the radial distance and axial coordinate, respectively; $A= 2 \times \frac{1}{2\pi s_0^2} \sqrt{\frac{bc}{\pi^2 }}$, $b,c$ and $s_0$ are constants that depend on the model potential.
In analogy to \Eq{ExA}, 
\begin{align}
	E_x^{\rm LDA}[\rho_{\text{mol}}] &= -C_x \int \rho_{\rm mol}^{4/3}d\br \approx -\frac{2\pi C_xA^{4/3}}{\hbar^{4/3}}\int_{-\infty}^\infty dz\int_0^\infty sds\exp\Big[ -\frac{4b}{3\hbar}z^2 - \frac{4c}{3\hbar}(s-s_0)^2 \Big]. \label{ExM1}
\end{align}
Here 
\begin{align}
	&\int_{-\infty}^\infty dz\exp\Big[ - \frac{4b}{3\hbar}z^2 \Big] = \sqrt{\frac{3\pi\hbar}{4b}}, \nl
	&\int_0^\infty sds \exp\Big[ - \frac{4c}{3\hbar}(s-s_0)^2 \Big] \approx s_0 \sqrt{\frac{3\pi\hbar}{4c}}.  \label{ExM2}
\end{align}
Therefore, 
\begin{align}
	E_x^{\rm LDA}[\rho_{\rm mol}] \approx  -\mO\Big(\hbar^{-\frac{1}{3}}\Big). 
\end{align}
For the correlation energy, one can numerically verify that the asymptotic leading term is also $\mO(\hbar^{-1/3})$ by substituting \Eq{rhoM} into the VWN5 functional.

Moreover, we can show that the Hartree term for diatomic molecules also diverges in the semiclassical limit. We prove this by calculating the integral in the $\bk$ space. The Fourier transform of $\rho_{\rm mol}$ can be calculated using the similar technique of \Eq{ExM1}--\eqref{ExM2}, giving 
\begin{align}
	\tilde{\rho}(\bk) = & \frac{1}{(2\pi)^{3/2}}\int \rho_{\rm mol}(\br) e^{i\bk\cdot\br} d\br 
	\approx \frac{2}{(2\pi)^{3/2}} J_0(s_0 k_s) \exp\Big( - \frac{\hbar}{4 b} k_z^2  \Big),  
\end{align}
where $k_s$ and $k_z$ are the radial distance and axial coordinate in the reciprocal space, $J_0$ is the Bessel function of the first kind.
The Hartree term is given by
\begin{align}
	J = & \frac{1}{2} \int \frac{4\pi}{k^2} |\tilde{\rho}(\bk)|^2 d\bk \approx \frac{4}{\pi} \int_{0}^{+\infty} dk_z \cdot \exp\Big( - \frac{\hbar}{2b} k_z^2 \Big) \int_0^\infty k_s  dk_s \cdot [J_0(s_0 k_s)]^2  \frac{1}{k_s^2 + k_z^2} . \label{int-J}
\end{align}
Here we have used the fact that the integrand is an even function of $k_z$. Moreover, one can show that integration over the large-$k_z$ region makes $J$ blow up when $\hbar\to 0$. To analyze the leading order divergence, it suffices to replace the lower limit of 0 by 1 (to avoid singularity in the subsequent steps) in the integration over $k_z$ and analyze the asymptotic behavior of the second integral in \Eq{int-J} as a function of $k_z$. In particular, one can show that
\begin{align}
	\int_0^\infty dk_s \cdot k_s [J_0(s_0 k_s)]^2 \frac{1}{k_s^2 + k_z^2} \approx  \frac{1}{2k_z s_0} .  \label{approx-integral}
\end{align}
In deriving \Eq{approx-integral}, we have used $k_s[J_0(s_0 k_s)]^2 \approx \frac{1}{\pi s_0} [1 + \sin(2 s_0 k_s)]$. Here $\sin(2 s_0 k_s)$ is an oscillatory function of $k_s$, giving negligible contribution to the integral when multiplied by $\frac{1}{k_s^2 + k_z^2}$; replacing  $k_s[J_0(s_0 k_s)]^2$ by $k_s[J_0(s_0 k_s)]^2$ yields the final result of $\frac{1}{2k_z s_0}$.  
Using the above arguments in \Eq{int-J}, we have
\begin{align}
	J \approx & \frac{4}{\pi} \int_{1}^{+\infty} dk_z \cdot \exp\Big( - \frac{\hbar}{2b} k_z^2 \Big) \frac{1}{2k_z s_0}. \label{J}
\end{align}
The approximate formula for $J$ is $\hbar$-dependent. To better understand its $\hbar$-dependence, particularly for small $\hbar$, we evaluate
\begin{align}
	\frac{dJ}{d\hbar} = - \frac{4}{\pi} \int_{1}^{+\infty} dk_z \cdot \exp\Big( - \frac{\hbar}{2b} k_z^2 \Big) \frac{k_z}{4 b s_0} = - \frac{1}{\pi s_0}\frac{1}{\hbar} e^{-\frac{\hbar}{2b}} = -\frac{1}{\pi s_0}\frac{1}{\hbar} + \mO(1).
\end{align}
This suggests that 
\begin{align}
	J = \frac{1}{\pi s_0}|\ln \hbar| + \mO(1),
\end{align}
which has been verified by numerical results.

\section{Supplementary figure showing our effective potential for diatomic molecular model}

\begin{figure}[htpb]
	\begin{center}
		\includegraphics[width=0.8\linewidth]{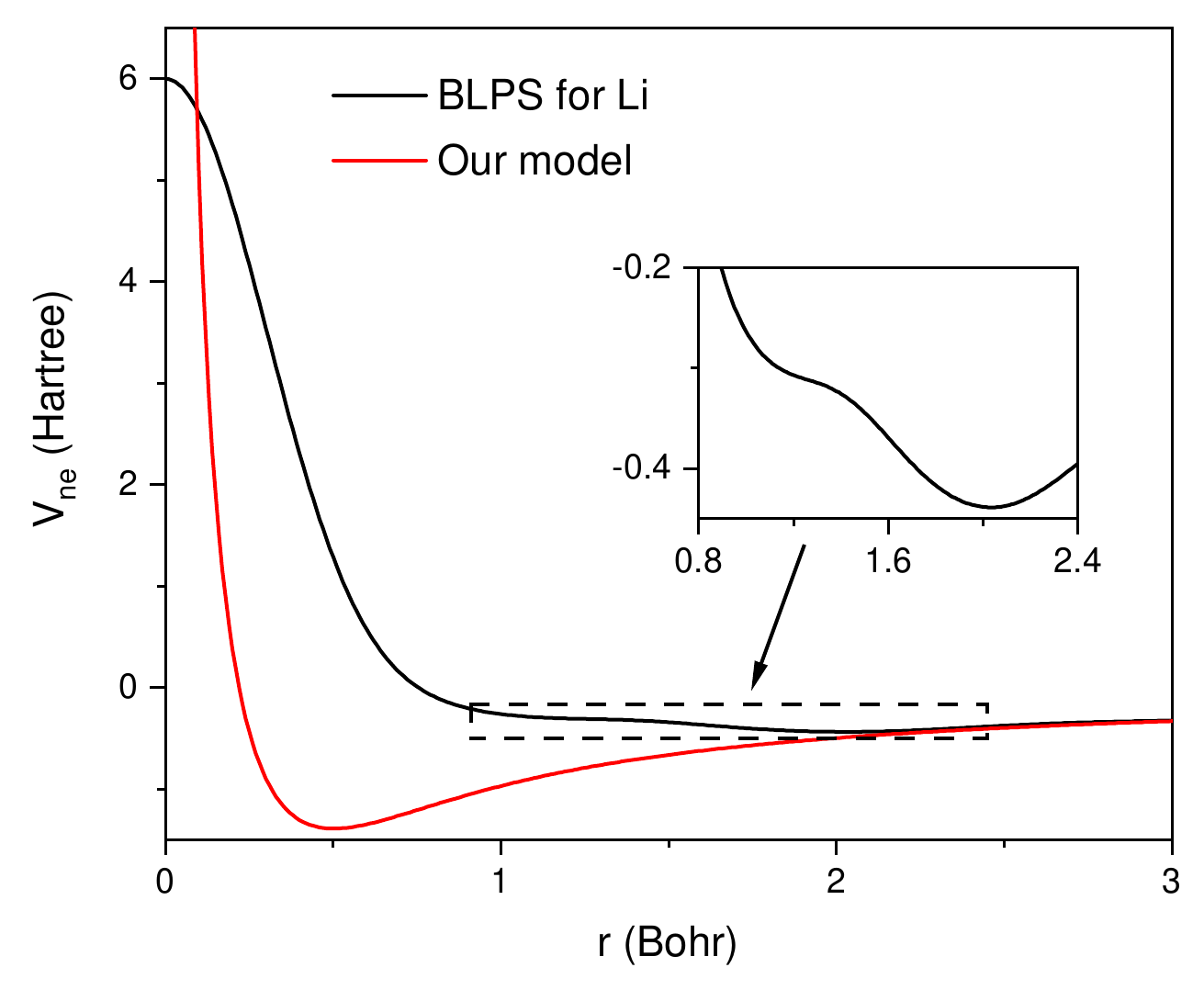}
	\end{center}
	\caption{
		Our self-designed effective potential as a function of electron-nuclear distance in comparison with the BLPS potential for lithium.
		The inset shows an enlarged plot of the intermediate region of BLPS to better visualize the fine details.
	} \label{VneMol}
\end{figure}

In \Fig{VneMol}, we compare our self-designed effective electron-nuclear potential for simulating single-bond stretching of a diatomic molecule with the bulk-derived local pseudopotential (BLPS) \cite{Huang087109} for Li atom.
Our potential has the following analytic form,
\begin{align}
	V_{\rm reff}(r) = -\frac{e^{\beta r} - 1}{e^{\beta r} + 1} \frac{1}{r} + \frac{2}{e^{\beta r}+1}\frac{1}{r}, 
\end{align} 
where $\beta=5$ is a adjustable parameter.
Such a form bridges the short-range repulsion and long-range attraction of the BLPS through a simple Fermi-Dirac function, and smooths out the fine details in the intermediate region in order to simplify the physical picture. This suffices for our need of proving concept. 

~\\
~\\

{\bf References}

\bibliographystyle{achemso}
\bibliography{Ref}